\documentclass[11pt]{article}
\usepackage{amsfonts}
\usepackage[english]{babel}
\usepackage{graphicx}
\usepackage{epstopdf}
\usepackage{epsfig}
\usepackage{amssymb}
\usepackage{setspace}
\usepackage{caption}
\usepackage{subcaption}
\usepackage{amsfonts}
\usepackage{color}
\usepackage{amsmath}
\usepackage{float}
\usepackage{comment}
\usepackage{amsmath}
\numberwithin{equation}{section}
\newcommand {\be}{\begin{equation}}
\newcommand {\ee}{\end{equation}}
\newcommand {\bea}{\begin{array}}
	
	\newcommand {\eea}{\end{array}}

\begin{document}

\begin{titlepage}
	\vspace{1 cm}
	\begin{center}
	{\Large \bf {Equatorial Circular Motion of Charged Test Particles in a Weakly Magnetized Taub--NUT Background}}\\
	\end{center}
	\vspace{1cm}
	\begin{center}
	\renewcommand{\thefootnote}{\fnsymbol{footnote}} B.J. Bansawang{\footnote{bansawang@fmipa.unhas.ac.id}} and Tasrief Surungan{\footnote{tasrief@unhas.ac.id}}\\
		  \vspace*{1cm}Department of Physics, Faculty of
		  Mathematics and Natural Sciences, Hasanuddin University,\\
		Makassar, South Sulawesi 90245, Indonesia
		\renewcommand{\thefootnote}{\arabic{footnote}} \vspace*{0.25cm}
		\\
	\end{center}
\begin{abstract}
We study circular motion of charged test particles on the equatorial slice of a
Taub--NUT black hole with Manko--Ruiz parameter $C$, immersed in a weak external
magnetic field introduced via Wald's prescription. Because the Taub--NUT metric is
not reflection-symmetric about the equator once $l\neq 0$, generic charged orbits
lie on cones $x=\cos\theta\neq 0$ rather than on the equatorial plane. We therefore
analyse \emph{constrained} circular orbits obtained by imposing $x=\dot x=0$, and
we exhibit in closed form the residual angular constraint that a fully
self-consistent orbit would have to satisfy. Within this scope we derive the
circularity and marginal-stability conditions and study how $B$ and $C$ shift the
ISCO radius for prograde and retrograde branches. Increasing $B$ monotonically
decreases $r_{\mathrm{ISCO}}$; the sign of the particle charge splits the two
branches, with the ordering reversed between prograde and retrograde motion; and
$C$ contributes only subleading corrections. The extension to self-consistent
conical orbits is the natural direction for follow-up work.
	\end{abstract}
\end{titlepage}\onecolumn
\bigskip

\section{Introduction}
\label{sec:intro}
The Taub--NUT spacetime is one of the most intriguing vacuum solutions of Einstein's
gravitational field equations. It contains an additional parameter --- the NUT charge ---
first introduced by Taub \cite{Taub:1950ez} and later generalized by Newman, Tamburino,
and Unti \cite{Newman:1963yy}. This parameter induces nontrivial gravitomagnetic
properties that are absent from both the Schwarzschild and Reissner--Nordstr\"om
solutions, and it is associated with coordinate singularities known as Misner strings
\cite{Misner:1963fr}. The physical interpretation of these singularities has been a
long-standing point of discussion: Bonnor \cite{Bonnor:1969} interpreted them as
semi-infinite massless sources of angular momentum; Cl\'ement, Gal'tsov, and Guenouche
\cite{Clement:2015cxa} subsequently argued that the corresponding closed timelike
curves are physically harmless and that Taub--NUT geometries should not be discarded
on this basis; and a more modern thermodynamic treatment of the Lorentzian Taub--NUT
solution was developed in \cite{Hennigar:2019ive}. A modern gauge formulation of the
Misner-string structure, introduced by Manko and Ruiz \cite{Manko:2005nm}, parameterizes
the distribution of these singularities through a constant $C$, with $C=-1$ placing the
string at the north pole, $C=+1$ at the south pole, and $C=0$ distributing the string
symmetrically between the two poles. Geodesic motion in the Taub--NUT background has
been studied in detail by Kagramanova, Kunz, Hackmann, and L\"ammerzahl
\cite{Kagramanova:2010bk}.

Research on magnetized black holes has attracted increasing attention in light of
observational evidence for strong magnetic fields in the vicinity of supermassive black
holes. Polarimetric imaging by the Event Horizon Telescope Collaboration of the
supermassive black hole at the centre of M87
\cite{EHT:2021ring,EHT:2021magnetic} provides direct evidence for an ordered
horizon-scale magnetic field, while measurements of Faraday rotation toward the
magnetar PSR\,J1745--2900 in the Galactic Centre have constrained the magnetic field
near Sagittarius~A$^{\ast}$ \cite{Eatough:2013qua}. These observations motivate detailed
theoretical investigations of charged-particle dynamics in magnetized black-hole
spacetimes.

The standard framework for introducing an external electromagnetic field without
deforming the spacetime geometry is Wald's prescription \cite{Wald:1974np}, which
exploits the Killing symmetries of the background. Charged-particle dynamics in
magnetized Schwarzschild and Kerr black-hole geometries have been analysed extensively
\cite{Aliev:1989,Frolov:2010mi,Kolos:2015iva,Tursunov:2016xtj,Kolos:2017ojf}, with
particular emphasis on the displacement of the innermost stable circular orbit (ISCO)
by the Lorentz coupling and on the resulting epicyclic frequencies as candidate
explanations of the high-frequency quasi-periodic oscillations observed in microquasars.
The combined effect of the magnetic field and the NUT charge on charged-particle
collisions and ISCO has been investigated in
\cite{Abdujabbarov:2012bn,Shaymatov:2021nff}. The study of magnetized variants of
Taub--NUT and related geometries within Einstein--Maxwell theory has been extensively
developed by Siahaan and collaborators
\cite{Siahaan:2021ypk,Siahaan:2023bhc,Siahaan:2021ags,Siahaan:2021uqo,Ghezelbash:2021lcf,Siahaan:2024fbh,Siahaan:2024ioa,Ghezelbash:2021xvc},
and a comprehensive treatment of exact solutions of this family is given in the
monograph of Griffiths and Podolsk\'y \cite{griffiths}.

In this paper we extend this programme to the dynamics of charged test particles in a
weakly magnetized Taub--NUT background, with explicit dependence on the Manko--Ruiz
parameter $C$. Our work differs from the closest prior analyses
\cite{Siahaan:2021ypk,Siahaan:2023bhc,Abdujabbarov:2012bn} in three respects: (i) we
provide a systematic, gauge-aware analysis of $r_{\mathrm{ISCO}}$ as a function of both
the field strength $B$ and the Misner-string parameter $C$ for prograde and retrograde
branches; (ii) we make the scope of our analysis fully explicit by examining whether
$x=\dot x=0$ is dynamically self-consistent in this background, exhibiting the residual
angular constraint that obstructs identification of the equatorial slice with a true
orbit family in the generic case, and treating our results as those of constrained
circular motion on the imposed slice; and (iii) we derive the marginal-stability
condition in a clean, manifestly self-consistent form by treating the effective
conserved quantities $\mathcal{E}_{\mathrm{eff}}$ and $\mathcal{L}_{\mathrm{eff}}$ as
$r$-dependent combinations from the outset.

The remainder of the paper is organised as follows. Section~\ref{sec:background}
introduces the weakly magnetized Taub--NUT spacetime and the Wald electromagnetic
potential. Section~\ref{sec:circular} sets up the charged-particle Lagrangian, examines
the angular Euler--Lagrange equation and the residual angular constraint at $x=0$ that
generic charged orbits do not automatically satisfy, and derives the circularity and
marginal-stability conditions in a clean form for circular motion on the imposed
equatorial slice. Section~\ref{sec:emforce} analyses the Lorentz force, the
magnetic-field components measured by a locally non-rotating observer, and the
dependence of the constrained ISCO radius and conserved quantities on $B$ and $C$ for
both prograde and retrograde branches. We close with a brief outlook on orbital
stability via Lyapunov exponents and summarise our findings, including the limitations
of the equatorial-slice approximation, in Section~\ref{sec:conclusion}. Throughout the
paper we use natural units with $G=c=1$ and the metric signature $(-,+,+,+)$.

\section{Weakly Magnetized Taub--NUT Spacetime}\label{sec:background}
\subsection{Background geometry}

We consider the Lorentzian Taub--NUT metric, a stationary axisymmetric vacuum solution
to Einstein's field equations. Working in coordinates $(t,r,x,\phi)$ with $x = \cos\theta$,
the line element reads
\begin{equation}
	ds^2 = -\frac{\Delta}{\Sigma} \bigl(dt -2l(x+C)\, d\phi \bigr)^2 + \frac{\Sigma}{\Delta}dr^2 + \Sigma\left(\frac{dx^2}{1-x^2} + (1-x^2)\, d\phi^2\right),
\end{equation}
where
\begin{equation}
	\Delta = r^2-2Mr-l^2,\qquad \Sigma = r^2+l^2.
\end{equation}
Here $M$ is the gravitating mass, $l$ is the NUT parameter (gravitomagnetic charge), and
the coordinate $x$ ranges over $[-1,1]$. The Manko--Ruiz constant $C$ governs the
distribution of the Misner-string singularities: $C=-1$ places the string at the north
pole ($x=1$), $C=+1$ at the south pole ($x=-1$), and $C=0$ distributes the strings
symmetrically between the poles.

\subsection{Wald electromagnetic potential}
To introduce a magnetic field without backreacting on the geometry (test-field
approximation), we follow Wald's prescription \cite{Wald:1974np}. For a stationary,
axisymmetric spacetime with Killing vectors $\xi_{(t)}=\partial_t$ and
$\xi_{(\phi)}=\partial_\phi$, a uniform magnetic field aligned with the symmetry axis
is encoded by the Killing four-potential
\begin{equation}\label{eq:waldA}
	A_{\mu}=B\,\xi_{(\phi)\mu}=B\,g_{\mu\phi}.
\end{equation}
We adopt this convention throughout: the symbol $B$ denotes the parameter of the
Killing potential rather than the asymptotic field strength measured by a static
observer at infinity, which differs from $B$ by an overall numerical factor depending
on the spacetime. Since all of our results depend on $B$ only through its appearance
in $A_\mu$, the physics is unaffected; this fixes a single, internally consistent
convention.

The non-vanishing components in the $(t,r,x,\phi)$ basis are
\begin{align}
	A_t &= B\,g_{t\phi}=2Bl\frac{\Delta}{\Sigma}(x+C), \label{eq:At}\\
	A_\phi &= B\,g_{\phi\phi}=B\left[\Sigma(1-x^2)-4l^2 \frac{\Delta}{\Sigma}(x+C)^2\right].\label{eq:Aphi}
\end{align}
The non-zero $A_t$ arises from the $t$--$\phi$ mixing induced by the NUT parameter:
unlike the Kerr case, this contribution survives in the static limit. In the symmetric
gauge $C=0$ evaluated at the equatorial plane $x=0$, one has $A_t=0$, so only $A_\phi$
couples to the particle charge there.

\section{Circular motion of charged particles}\label{sec:circular}
\subsection{Conserved quantities and effective potential}
The Lagrangian for a massive charged test particle of specific charge $e\equiv q/\mu$,
where $q$ and $\mu$ are the particle charge and rest mass respectively, is
\begin{equation}
{\cal L} = \tfrac{1}{2}g_{\mu\nu}\,\dot x^\mu\dot x^\nu + e A_\mu \dot x^\mu,
\end{equation}
where overdots denote differentiation with respect to proper time $\tau$. Timelike
normalization requires $g_{\mu\nu}\dot x^\mu\dot x^\nu=-1$. The stationarity and
axisymmetry of the background yield two conserved quantities:
\begin{align}
-E &\equiv \frac{\partial {\cal L}}{\partial \dot t} = g_{tt}\dot t + g_{t\phi}\dot\phi + eA_t,\\
L &\equiv \frac{\partial {\cal L}}{\partial \dot\phi} = g_{t\phi}\dot t + g_{\phi\phi}\dot\phi + eA_\phi.
\end{align}
Inverting these relations gives the $t$ and $\phi$ components of the four-velocity:
\begin{align}
\dot t &= \frac{g_{t\phi}\,(L - eA_\phi) + g_{\phi\phi}\,(E + eA_t)}{\mathcal{D}},\\
-\dot\phi &= \frac{g_{tt}\,(L - eA_\phi) + g_{t\phi}\,(E + eA_t)}{\mathcal{D}},
\end{align}
where $\mathcal{D}=g_{t\phi}^2-g_{tt}g_{\phi\phi}$ is the determinant of the $t$--$\phi$
block of the metric. It is convenient to define the effective conserved quantities
\begin{equation}\label{eq:effEL}
\mathcal{E}_{\mathrm{eff}}(r,x)\equiv E+eA_t(r,x),\qquad \mathcal{L}_{\mathrm{eff}}(r,x)\equiv L-eA_\phi(r,x),
\end{equation}
in terms of which the four-velocity components take the compact form
\begin{equation}
\dot t = \frac{g_{t\phi}\,\mathcal{L}_{\mathrm{eff}} + g_{\phi\phi}\,\mathcal{E}_{\mathrm{eff}}}{\mathcal{D}},\qquad
-\dot\phi = \frac{g_{tt}\,\mathcal{L}_{\mathrm{eff}} + g_{t\phi}\,\mathcal{E}_{\mathrm{eff}}}{\mathcal{D}}.
\end{equation}
Substituting into the normalization condition
\begin{equation}
g_{tt}\dot t^2 + g_{rr}\dot r^2 + g_{xx}\dot x^2 + 2g_{t\phi}\dot t\dot\phi + g_{\phi\phi}\dot\phi^2 = -1
\end{equation}
yields the radial equation
\begin{equation}\label{eq:radialgeneral}
g_{rr}\dot r^2 + g_{xx}\dot x^2 = -1 + \frac{1}{\mathcal{D}}\Big[g_{tt}\,\mathcal{L}_{\mathrm{eff}}^2 + 2g_{t\phi}\,\mathcal{L}_{\mathrm{eff}}\,\mathcal{E}_{\mathrm{eff}} + g_{\phi\phi}\,\mathcal{E}_{\mathrm{eff}}^2\Big],
\end{equation}
where the $(t,\phi)$-block contribution has been evaluated using
$g_{tt}\dot t + g_{t\phi}\dot\phi = -\mathcal{E}_{\mathrm{eff}}$ and
$g_{t\phi}\dot t + g_{\phi\phi}\dot\phi = \mathcal{L}_{\mathrm{eff}}$ together with
$\mathcal{D} = g_{t\phi}^2 - g_{tt}g_{\phi\phi}$.

\subsection{The angular constraint at $x=0$ and the scope of equatorial circular motion}\label{sec:eqconsistency}

Unlike Schwarzschild and Kerr, the Taub--NUT metric is \emph{not} reflection-symmetric
about $x=0$ once $l\neq 0$: the term $-2l(x+C)\,d\phi$ in (2.1) is odd in $x$ for
$C=0$ and asymmetric for $C\neq 0$. Consequently, $x=0,\ \dot x=0$ is not a fixed
point of an isometry of the background, and one must check whether it is dynamically
consistent --- i.e.\ whether $\ddot x|_{x=0,\,\dot x=0}=0$ --- before treating it as
an orbit family. The Euler--Lagrange equation for $x$ is
\begin{equation}
\frac{d}{d\tau}\!\left(g_{xx}\dot x\right) - \tfrac{1}{2}\partial_x g_{\mu\nu}\dot x^\mu\dot x^\nu - e\bigl(\partial_x A_\mu - \partial_\mu A_x\bigr)\dot x^\mu = 0.
\end{equation}
With $A_x=0$ in our gauge and on the slice $x=\dot x=0$, the first term vanishes and
the surviving contributions are the partial $x$-derivatives of $g_{\mu\nu}$ and $A_\mu$
evaluated at $x=0$, contracted with $\dot t^2$, $\dot t\dot\phi$, $\dot\phi^2$, $\dot
t$, and $\dot\phi$. A direct computation from (2.1)--(2.3) and (\ref{eq:At})--(\ref{eq:Aphi})
yields
\begin{align}
\partial_x g_{tt}\big|_{x=0} &= 0,&
\partial_x g_{t\phi}\big|_{x=0} &= \frac{2l\Delta}{\Sigma},\\
\partial_x g_{\phi\phi}\big|_{x=0} &= -\frac{8l^2 C\Delta}{\Sigma},&
\partial_x g_{xx}\big|_{x=0} &= 0,\\
\partial_x A_t\big|_{x=0} &= \frac{2Bl\Delta}{\Sigma},&
\partial_x A_\phi\big|_{x=0} &= -\frac{8Bl^2 C\Delta}{\Sigma}.
\end{align}
Note in particular that $\partial_x g_{\phi\phi}|_{x=0}$ does not vanish for $C\neq 0$
--- a correction overlooked in earlier discussions of this slice. Substituting into
the Euler--Lagrange equation,
\begin{equation*}
-\tfrac{1}{2}\partial_x g_{tt}\dot t^2 -\partial_x g_{t\phi}\dot t\dot\phi -\tfrac{1}{2}\partial_x g_{\phi\phi}\dot\phi^2 -e\,\partial_x A_t\dot t -e\,\partial_x A_\phi \dot\phi = 0,
\end{equation*}
the condition $\ddot x|_{x=0}=0$ becomes
\begin{equation*}
-\frac{2l\Delta}{\Sigma}\dot t\dot\phi + \frac{4l^2 C\Delta}{\Sigma}\dot\phi^2 - \frac{2eBl\Delta}{\Sigma}\dot t + \frac{8eBl^2 C\Delta}{\Sigma}\dot\phi = 0.
\end{equation*}
After factoring the common prefactor $-2l\Delta/\Sigma$ (which is nonzero for
$l\neq 0$ and outside the horizon), this reduces to the single algebraic constraint
\begin{equation}\label{eq:angconstraint}
\dot t\,(\dot\phi + eB) \;=\; 2lC\,\dot\phi\,(\dot\phi + 2eB).
\end{equation}

Equation (\ref{eq:angconstraint}) is a fact, not an identity. For generic circular
orbits with $\dot t\neq 0$ and $\dot\phi\neq 0$, it is \emph{not} satisfied: in the
symmetric gauge $C=0$ it collapses to $\dot t(\dot\phi+eB)=0$, requiring the
non-generic locking $\dot\phi=-eB$ between the orbital frequency and the magnetic
coupling; for $C\neq 0$ it imposes a quadratic relation between $(\dot t,\dot\phi)$
and the parameters $(l,C,e,B)$. This reflects the well-known feature, established
already in the analytic geodesic studies of the Taub--NUT spacetime
\cite{Kagramanova:2010bk}, that generic timelike trajectories in this background
trace out cones $x=x_0\neq 0$ rather than lying on the $x=0$ plane. The presence of
the magnetic field does not remove this obstruction; it modifies it.

Two responses to this fact are possible. The first is to abandon the equatorial slice
and analyse self-consistent circular motion on cones $x=x_0$ where $x_0$ is determined
by $\ddot x|_{x_0,\dot x=0}=0$ jointly with the radial circularity conditions; this is
a nontrivial four-equation system in $(r_0,x_0,E,L)$ which we leave to a forthcoming
companion paper. The second is to adopt the equatorial slice as an externally imposed
slicing condition and to study \emph{constrained} circular motion --- orbits which
solve the radial circularity and stability conditions at $x=0$ but do not in general
satisfy (\ref{eq:angconstraint}). This is the route taken in the remainder of the
present paper. We adopt the second option not because it is fully satisfactory, but
because (a) it isolates the radial dynamics in a tractable form that has nonetheless
been the de facto framework of much of the related literature, (b) it provides a
controlled baseline against which the conical extension can be benchmarked, and (c)
the residual angular term in (\ref{eq:angconstraint}) is, in the test-field regime
$BM \ll 1$, a small correction whose effect on the radial ISCO can be estimated
perturbatively. Throughout the rest of the paper, ``ISCO'' should therefore be read
as ``constrained ISCO on the imposed equatorial slice'', and the figures that follow
should be interpreted accordingly.

\subsection{Effective radial potential on the imposed equatorial slice}

We now restrict the radial dynamics to the imposed equatorial slice $x=\dot x=0$,
treating the angular constraint (\ref{eq:angconstraint}) as discussed above. Setting
$x=\dot x=0$ in (\ref{eq:radialgeneral}) and defining the effective potential
$V_{\mathrm{eff}}\equiv \dot r^2$, this takes the compact form
\begin{equation}\label{eq:Veff}
V_{\mathrm{eff}}(r) = -\frac{1}{g_{rr}} + \frac{1}{g_{rr}\,\mathcal{D}}\Big[g_{tt}\,\mathcal{L}_{\mathrm{eff}}^2 + 2g_{t\phi}\,\mathcal{L}_{\mathrm{eff}}\,\mathcal{E}_{\mathrm{eff}} + g_{\phi\phi}\,\mathcal{E}_{\mathrm{eff}}^2\Big]\Bigg|_{x=0},
\end{equation}
with the equatorial values of the metric and potential components readily obtained
from (2.1) and (\ref{eq:At})--(\ref{eq:Aphi}) at $x=0$:
\begin{equation}
g_{tt}|_{x=0}=-\frac{\Delta}{\Sigma},\qquad
g_{t\phi}|_{x=0}=\frac{2lC\Delta}{\Sigma},\qquad
g_{\phi\phi}|_{x=0}=\Sigma-\frac{4l^2C^2\Delta}{\Sigma},
\end{equation}
\begin{equation}
A_t|_{x=0}=\frac{2BlC\Delta}{\Sigma},\qquad
A_\phi|_{x=0}=B\Bigl(\Sigma-\frac{4l^2C^2\Delta}{\Sigma}\Bigr).
\end{equation}
Equation (\ref{eq:Veff}) is the form we use throughout the remainder of the paper.
The circularity condition $V_{\mathrm{eff}}(r_0)=0$ is equivalent to
$g_{tt}\mathcal{L}_{\mathrm{eff}}^2 + 2g_{t\phi}\mathcal{L}_{\mathrm{eff}}\mathcal{E}_{\mathrm{eff}} + g_{\phi\phi}\mathcal{E}_{\mathrm{eff}}^2 = \mathcal{D}$,
which matches the definition of $N(r)$ used in
Section~\ref{sec:circularityISCO} below. The form (\ref{eq:Veff}) is manifestly
consistent with the structure of the conservation laws and avoids the intermediate
algebraic ambiguities that arise from expanding $\mathcal{E}_{\mathrm{eff}}$ and
$\mathcal{L}_{\mathrm{eff}}$ before differentiation.

\subsection{Circularity and ISCO conditions}\label{sec:circularityISCO}

A circular orbit at radius $r=r_0$ satisfies $V_{\mathrm{eff}}(r_0)=0$ and
$V'_{\mathrm{eff}}(r_0)=0$ simultaneously. Defining
\begin{equation}\label{eq:Ndef}
N(r) \equiv g_{tt}\,\mathcal{L}_{\mathrm{eff}}^2 + 2g_{t\phi}\,\mathcal{L}_{\mathrm{eff}}\,\mathcal{E}_{\mathrm{eff}} + g_{\phi\phi}\,\mathcal{E}_{\mathrm{eff}}^2 - \mathcal{D},
\end{equation}
the first circularity condition is
\begin{equation}\label{eq:N0}
N(r_0)=0.
\end{equation}
Because $\mathcal{E}_{\mathrm{eff}}$ and $\mathcal{L}_{\mathrm{eff}}$ depend on $r$
through $A_t(r)$ and $A_\phi(r)$, their $r$-derivatives are
\begin{equation}
\mathcal{E}'_{\mathrm{eff}} = e A'_t,\qquad \mathcal{L}'_{\mathrm{eff}} = -e A'_\phi.
\end{equation}
Differentiating (\ref{eq:Ndef}) once gives
\begin{equation}\label{eq:Nprime}
\begin{aligned}
N'(r) ={}& g'_{tt}\,\mathcal{L}_{\mathrm{eff}}^2 + 2g'_{t\phi}\,\mathcal{L}_{\mathrm{eff}}\,\mathcal{E}_{\mathrm{eff}} + g'_{\phi\phi}\,\mathcal{E}_{\mathrm{eff}}^2 - \mathcal{D}'\\
&-2e A'_\phi\bigl(g_{tt}\,\mathcal{L}_{\mathrm{eff}} + g_{t\phi}\,\mathcal{E}_{\mathrm{eff}}\bigr)
+2e A'_t\bigl(g_{t\phi}\,\mathcal{L}_{\mathrm{eff}} + g_{\phi\phi}\,\mathcal{E}_{\mathrm{eff}}\bigr).
\end{aligned}
\end{equation}
The second circularity condition is $N'(r_0)=0$. The marginal-stability (ISCO)
condition is $N''(r_0)=0$, which differentiating (\ref{eq:Nprime}) once more yields
\begin{equation}\label{eq:Ndoubleprime}
\begin{aligned}
N''(r) ={}& g''_{tt}\,\mathcal{L}_{\mathrm{eff}}^2 + 2g''_{t\phi}\,\mathcal{L}_{\mathrm{eff}}\,\mathcal{E}_{\mathrm{eff}} + g''_{\phi\phi}\,\mathcal{E}_{\mathrm{eff}}^2 - \mathcal{D}''\\
&- 4e A'_\phi\bigl(g'_{tt}\,\mathcal{L}_{\mathrm{eff}} + g'_{t\phi}\,\mathcal{E}_{\mathrm{eff}}\bigr)
+ 4e A'_t\bigl(g'_{t\phi}\,\mathcal{L}_{\mathrm{eff}} + g'_{\phi\phi}\,\mathcal{E}_{\mathrm{eff}}\bigr)\\
&-2e A''_\phi\bigl(g_{tt}\,\mathcal{L}_{\mathrm{eff}} + g_{t\phi}\,\mathcal{E}_{\mathrm{eff}}\bigr)
+2e A''_t\bigl(g_{t\phi}\,\mathcal{L}_{\mathrm{eff}} + g_{\phi\phi}\,\mathcal{E}_{\mathrm{eff}}\bigr)\\
&+2e^2\Bigl[g_{tt}\,(A'_\phi)^2 + g_{\phi\phi}\,(A'_t)^2 - 2g_{t\phi}\,A'_t A'_\phi\Bigr].
\end{aligned}
\end{equation}
The simultaneous solution of $N=N'=N''=0$ provides the triple
$\{E,L,r_{\mathrm{ISCO}}\}$ for any given parameter set $\{M,l,B,C,e\}$. In the
neutral limit $e\to 0$ these conditions reduce to the standard geodesic circularity
conditions in the Taub--NUT spacetime; in the further limit $l\to 0$ they reduce to
the magnetized Schwarzschild conditions, and in the additional limit $B\to 0$ to the
familiar Schwarzschild result $r_{\mathrm{ISCO}}=6M$. We have used these limits as
numerical consistency checks for our code.

\subsection{Effective potential: qualitative behaviour}
Figure~1 shows $V_{\mathrm{eff}}(r)$ on the equatorial plane for the three gauge
choices $C=0,\pm 1$. As expected in the test-field regime, the magnetic field deforms
the potential wells only mildly. Nevertheless, both the electromagnetic coupling and
the position of the Misner strings shift the depth and location of the minima of
$V_{\mathrm{eff}}=0$, thereby modifying the radii and stability properties of circular
orbits relative to the purely geodesic neutral case with $C=0$.
 \begin{figure}[ht!]
	\centering
	\begin{subfigure}[b]{0.45\textwidth}
		\includegraphics[width=\textwidth]{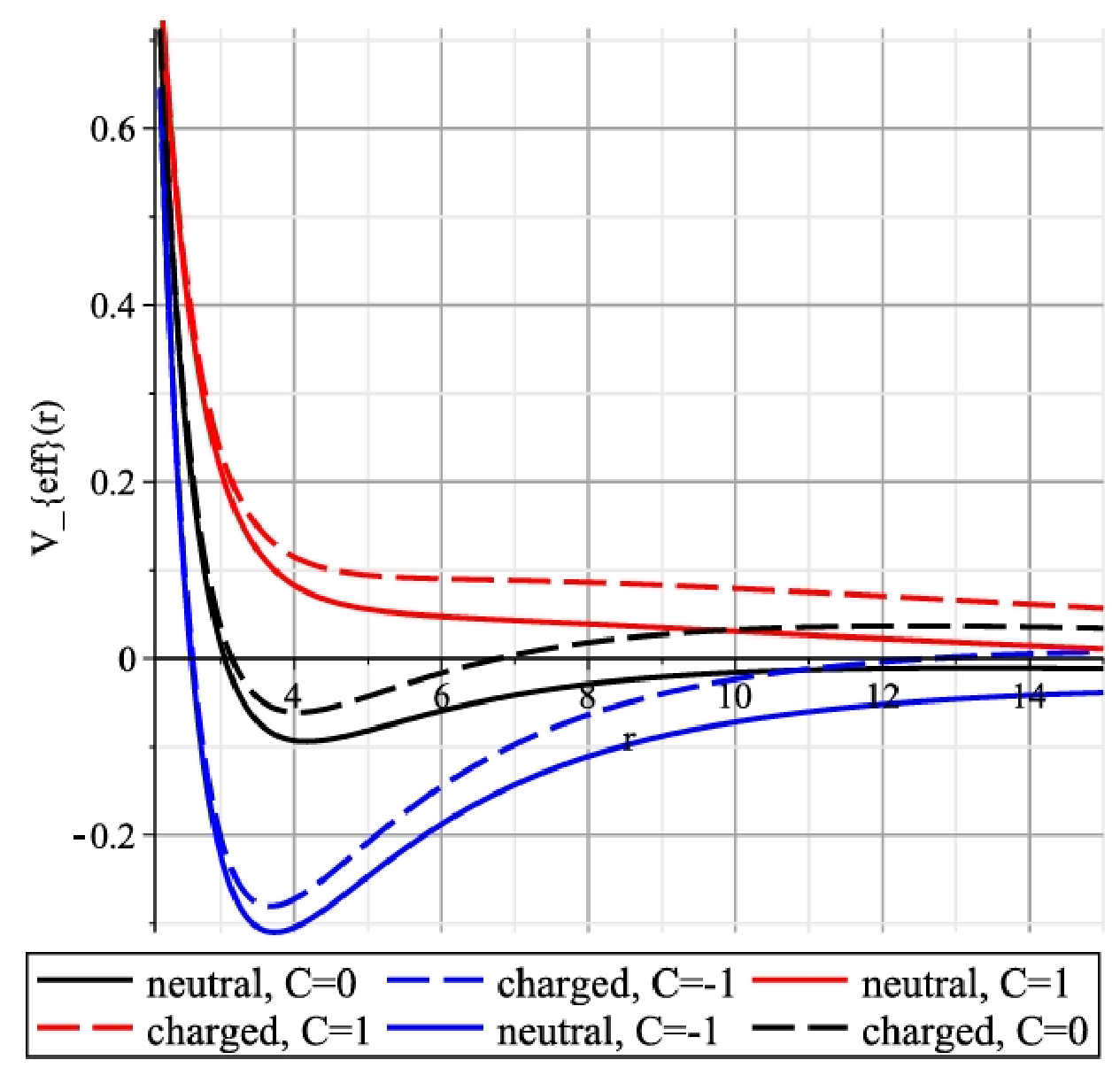}
		\label{fig:Veff(r)}
	\end{subfigure}
	\caption{Effective radial potential $V_{\rm eff}(r)$ for neutral ($e = 0$, solid curves) 
and charged ($e = 0.8$, dashed curves) massive test particles on the equatorial plane 
of the weakly magnetized Taub--NUT spacetime. Parameters: $M = 1$, $l = 0.4$, 
$BM = 0.01$, $(E,L) = (0.96, 4.2)$. Colors encode the Misner--string gauge: 
black $(C = 0)$, red $(C = +1$, string at south pole), blue $(C = -1$, 
string at north pole). Both the electromagnetic coupling and the string placement 
produce mild deformations of the potential well, shifting the location and depth 
of the circular-orbit minimum even in the weak-field regime.}
  \end{figure}
 
\section{Circular Orbits and Electromagnetic Force}\label{sec:emforce}
The circular motion of a charged particle around a magnetized Taub--NUT black hole
is governed by the interplay of gravity, the gravitomagnetic NUT effect, and the
Lorentz force due to the external magnetic field. The electromagnetic coupling
modifies both the conserved angular momentum and the effective energy, so that the
orbit is no longer purely geodesic.

\subsection{Lorentz force and observer-dependent fields}
The covariant Lorentz four-force on a charged particle is \cite{landau}
\begin{equation}\label{eq:lorentz}
f^\mu = \frac{q}{\mu}\,F^{\mu}{}_{\nu}\,u^{\nu},
\end{equation}
where $F_{\mu\nu}=\partial_\mu A_\nu - \partial_\nu A_\mu$ is the electromagnetic
field tensor and $u^\mu$ is the particle four-velocity. The decomposition of
$F^{\mu\nu}$ into electric and magnetic parts is in general observer-dependent:
because $A_t$ in (\ref{eq:At}) depends on both $r$ and $x$, the field tensor has
non-vanishing components $F_{rt}$ and $F_{xt}$ in addition to the magnetic-type
components $F_{r\phi}$ and $F_{x\phi}$, so for an arbitrary observer the
configuration is not purely magnetic. As noted in
Section~\ref{sec:coordfield} below, we will not undertake an explicit
orthonormal-tetrad decomposition of the measured fields in this paper; we work
throughout with the coordinate-component $A_\mu$ and $F_{\mu\nu}$, which is
sufficient for the radial ISCO analysis we pursue.

The radial component of the Lorentz force on the equatorial plane reads
\begin{equation}\label{eq:fr}
f^{r} = -\,\frac{eB}{2\Sigma}\Bigl[2r - \frac{4l^2C^2}{\Sigma^2}\bigl(4rl^2+2M(r^2-l^2)\bigr)\Bigr]\,\Bigl\{E\,g_{\phi\phi} + (L-eA_\phi)\,g_{t\phi}\Bigr\},
\end{equation}
which is plotted in Fig.~2 for representative values of $C$. The Lorentz force acts
perpendicular to the particle's direction of motion in the $(r,\phi)$ plane and
modifies the effective radial force, leading to the familiar Larmor / anti-Larmor
classification: orbits for which the Lorentz force is centripetal (Larmor-type)
admit smaller stable circular orbits than orbits for which it is centrifugal
(anti-Larmor-type). The classification depends on the joint sign of $eB$ and $L$, so
prograde and retrograde orbits at fixed sign of $e$ realize opposite cases.

\begin{figure}[ht!]
	\centering
\begin{subfigure}{0.45\textwidth}
    \centering
    \includegraphics[width=\textwidth]{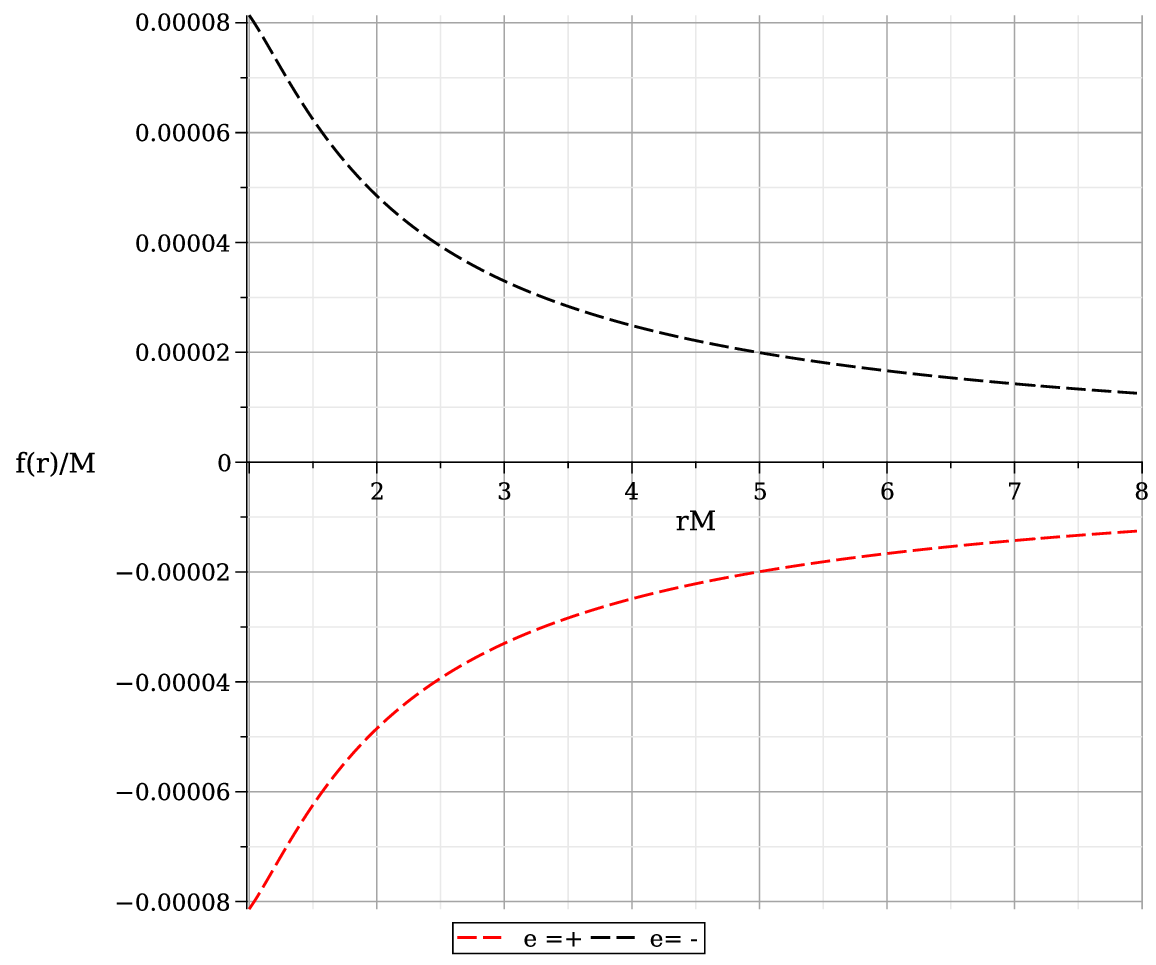}
    \caption{$C=1$}
    \label{fig:FrC1}
\end{subfigure}
\begin{subfigure}{0.45\textwidth}
    \centering
    \includegraphics[width=\textwidth]{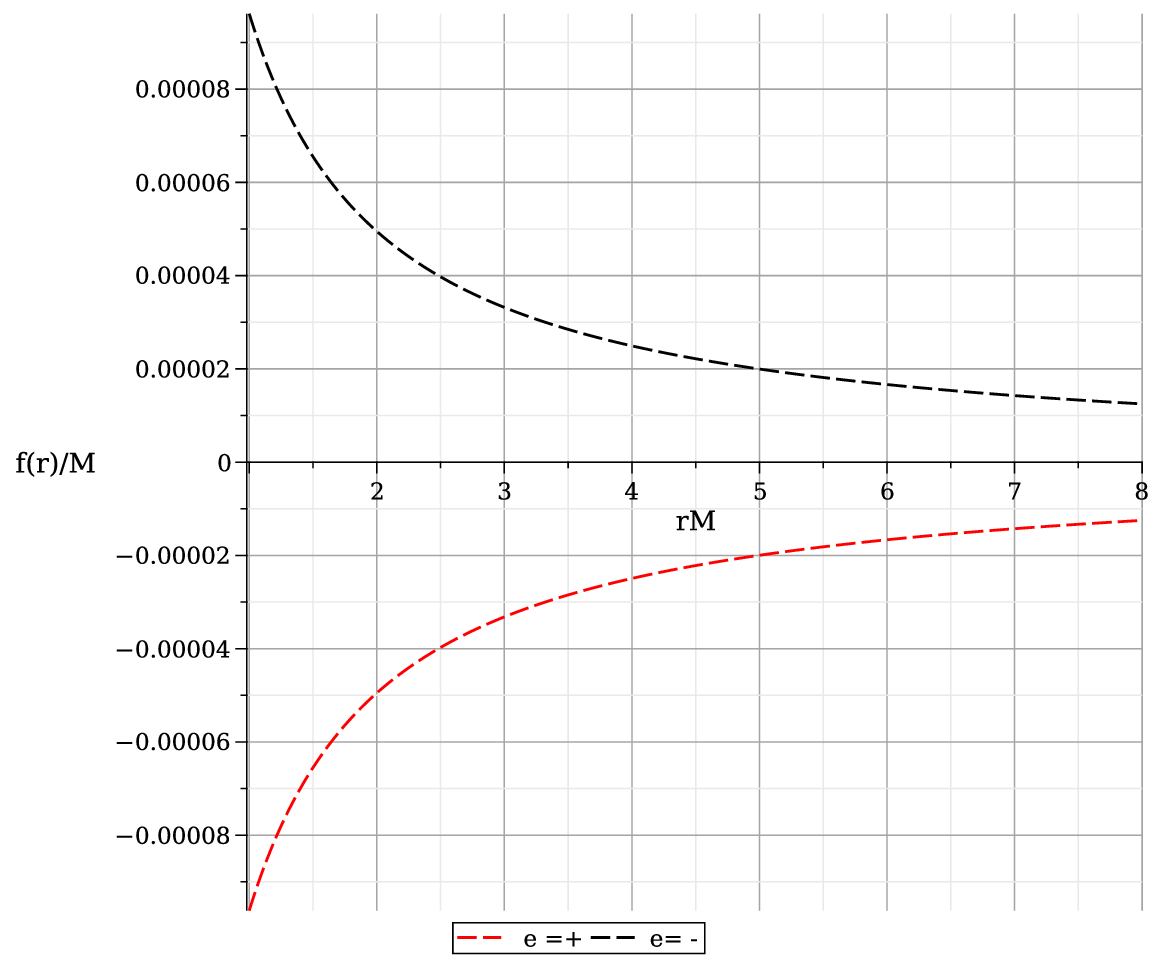}
    \caption{$C=0$}
    \label{fig:FrC0}
\end{subfigure}
\caption{Radial Lorentz force $f^{r}$ as a function of $r$ in the weakly magnetized
Taub--NUT spacetime with $l=0.2$. Panels (a)--(b) correspond to $C=1,0$. Red dashed
curves: positively charged particles ($e>0$); black dashed curves: negatively charged
particles ($e<0$).}
\end{figure}

\subsection{Coordinate-component electromagnetic field used downstream}
\label{sec:coordfield}

For the radial ISCO analysis that follows, only the coordinate components of $A_\mu$
in (\ref{eq:At})--(\ref{eq:Aphi}) and the corresponding $F_{\mu\nu}$ enter, through
the Lagrangian (\ref{eq:lorentz}) and the conserved quantities $E$ and $L$. We
therefore do not undertake the full orthonormal-tetrad decomposition of the
ZAMO-measured $E^{\hat\mu}$ and $B^{\hat\mu}$ components in the present work; an
unambiguous tetrad analysis of the measured fields, including the residual electric
component induced by the $r$- and $x$-dependence of $A_t$, is identified as a natural
follow-up direction. All references to ``measured'' or ``observer-dependent'' fields
in what follows should therefore be understood as referring to the coordinate
components of $F_{\mu\nu}$ rather than to a fully tetrad-decomposed observable.

\subsection{Prograde ISCO}
Throughout this and the following subsection, ``ISCO'' denotes the constrained
innermost stable circular orbit on the imposed equatorial slice $x=\dot x=0$, in the
sense made precise in Section~\ref{sec:eqconsistency}. The prograde branch refers to
orbits co-rotating with the effective dragging of the spacetime ($L>0$ in our
conventions). The analysis of the prograde constrained ISCO is important for
characterising how the geometry and the magnetic field jointly influence the radial
stability of orbits on this slice.

Figure~4 summarises the prograde ISCO behaviour. For all three gauge choices
$C=0,\pm 1$, the ISCO radius decreases monotonically as $B$ increases, reflecting the
fact that the Lorentz interaction pulls stable prograde orbits closer to the black
hole. The red and black curves ($e>0$ and $e<0$, respectively) are slightly split,
indicating a mild charge-sign dependence. The variation with $C$ induces only a
subleading shift of the curves, confirming that the dominant effect is the magnetic
field strength rather than the Misner-string placement.

\begin{figure}[H]
	\centering
\begin{subfigure}{0.328\textwidth}
    \centering
    \includegraphics[width=\textwidth]{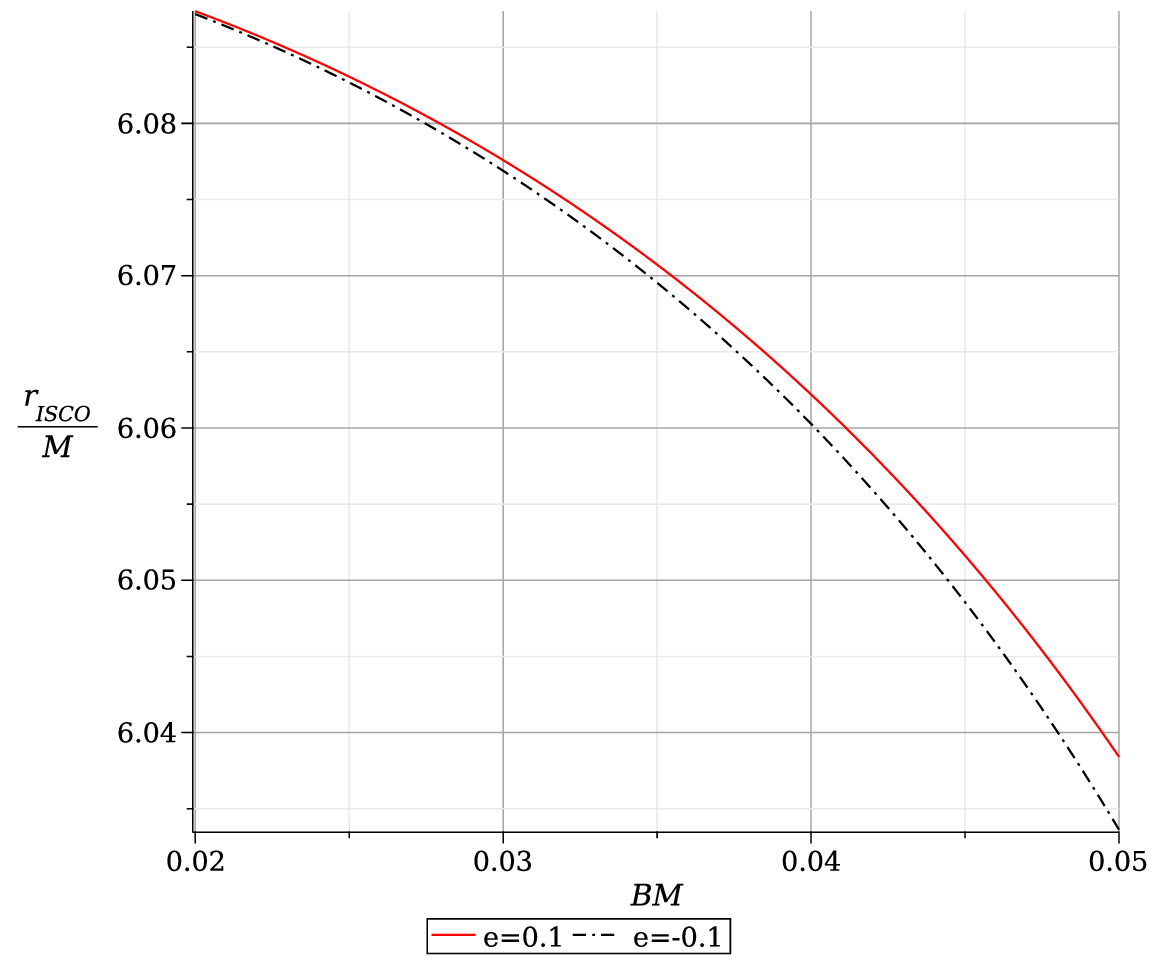}
    \caption{$C=-1$}
    \label{fig:progIsco_cminus1}
\end{subfigure}
\begin{subfigure}{0.328\textwidth}
    \centering
    \includegraphics[width=\textwidth]{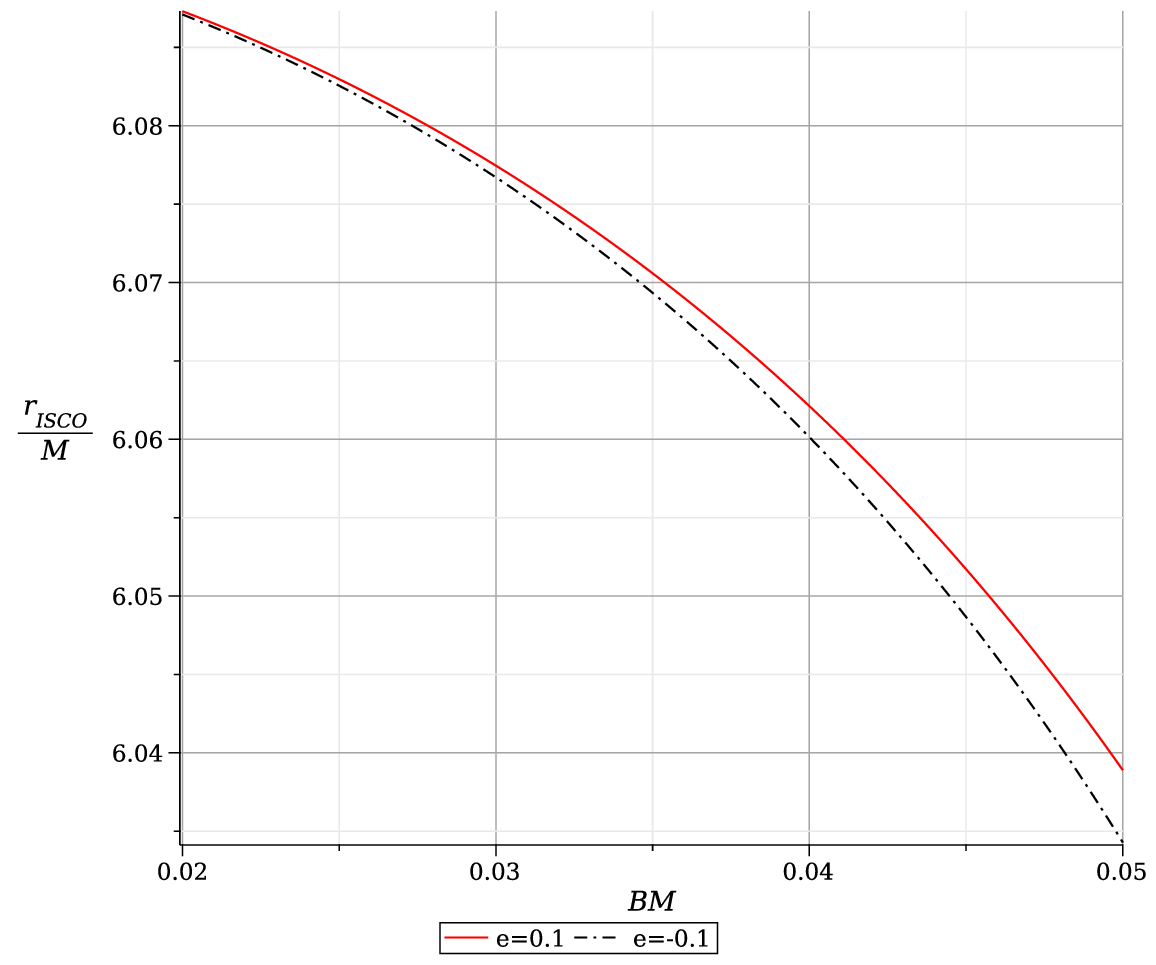}
    \caption{$C=0$}
    \label{fig:progIsco_c0}
\end{subfigure}
\begin{subfigure}{0.328\textwidth}
    \centering
    \includegraphics[width=\textwidth]{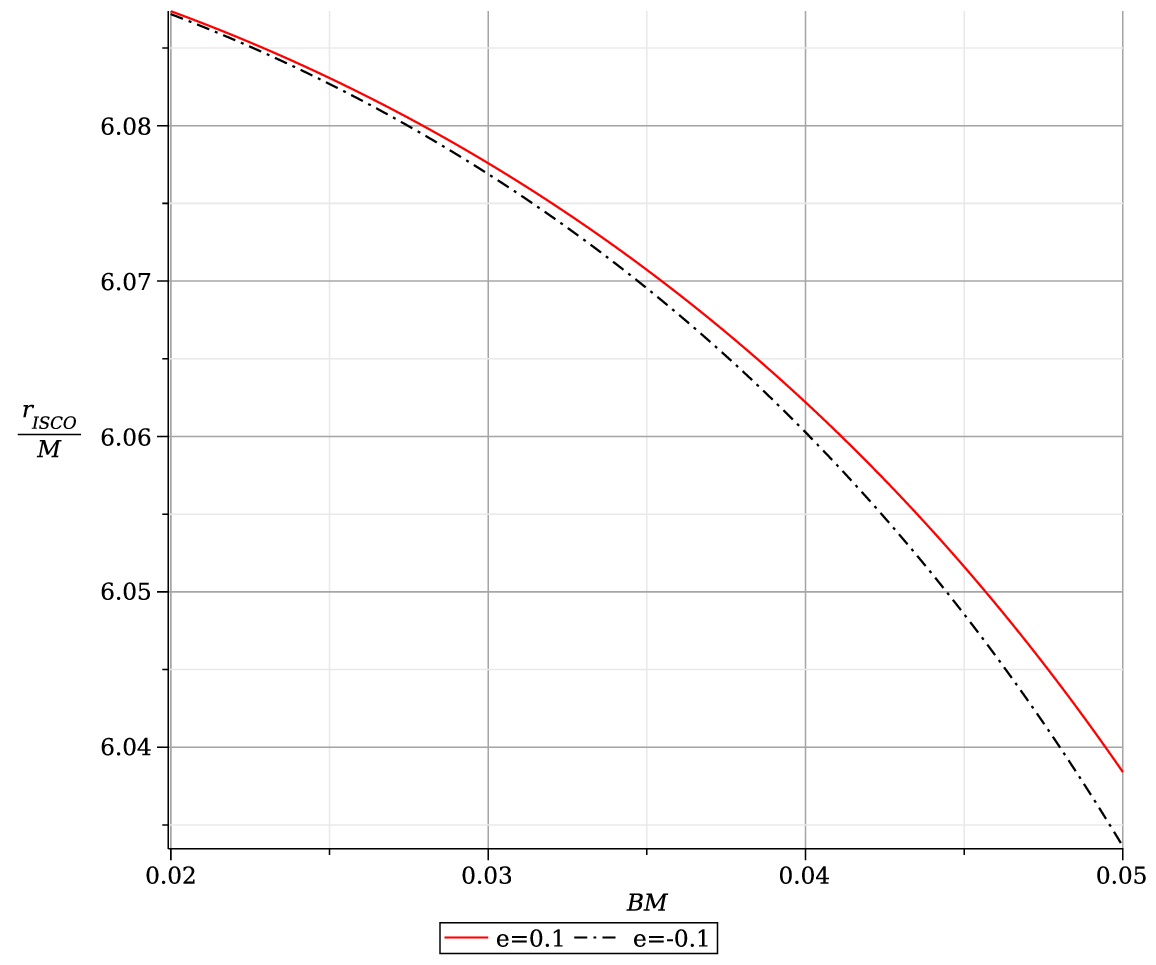}
    \caption{$C=1$}
    \label{fig:progIsco_c1}
\end{subfigure}

\caption{Prograde ISCO radius $r_{\mathrm{ISCO}}$ as a function of the external magnetic 
field strength $B$ in the weakly magnetized Taub--NUT spacetime with $l = 0.2$. 
Panels $(a)$--$(c)$ correspond to $C = -1, 0, +1$. 
Red solid curves: positively charged particles $(e > 0)$; 
black dash-dotted curves: negatively charged particles $(e < 0)$. 
Increasing $B$ monotonically decreases $r_{\mathrm{ISCO}}$; 
the charge sign produces a small but systematic splitting between the two branches, 
while the dependence on $C$ is subleading.}
\end{figure}

\subsection{Retrograde ISCO and the charge-sign asymmetry}
Figure~5 shows the retrograde ISCO behaviour. The qualitative trend mirrors the
prograde case: $r_{\mathrm{ISCO}}$ decreases monotonically with $B$ for both signs of
the charge. A notable asymmetry between the two cases emerges, however: for
retrograde motion the ISCO radius of positively charged particles decreases faster
with $B$ than that of negatively charged particles, whereas for prograde motion the
ordering is reversed.

This reversal admits a simple analytic explanation. The leading magnetic correction
to the ISCO radius is controlled by the linear coupling
$e\,B\,L_{\mathrm{ISCO}}^{(0)}$, where $L_{\mathrm{ISCO}}^{(0)}$ is the angular
momentum at the unperturbed ISCO. Switching the sign of $L$ (prograde
$\leftrightarrow$ retrograde) at fixed $eB$ flips the sign of this coupling and
therefore swaps which charge sign experiences the centripetal Larmor enhancement
versus the centrifugal anti-Larmor reduction. Concretely, denoting by
$\delta r_{\mathrm{ISCO}}^{\pm}$ the leading correction for $e=\pm|e|$ at fixed
prograde branch, one has
\begin{equation}
\delta r_{\mathrm{ISCO}}^{+}(L>0) \,=\, \delta r_{\mathrm{ISCO}}^{-}(L<0),\qquad
\delta r_{\mathrm{ISCO}}^{-}(L>0) \,=\, \delta r_{\mathrm{ISCO}}^{+}(L<0),
\end{equation}
to leading order in $B$. The crossing visible between Figs.~4 and 5 is exactly this
$L\to -L$ relabelling of the charge-sign branches.

  \begin{figure}[ht!]
	\centering
\begin{subfigure}{0.32\textwidth}
    \centering
    \includegraphics[width=\textwidth]{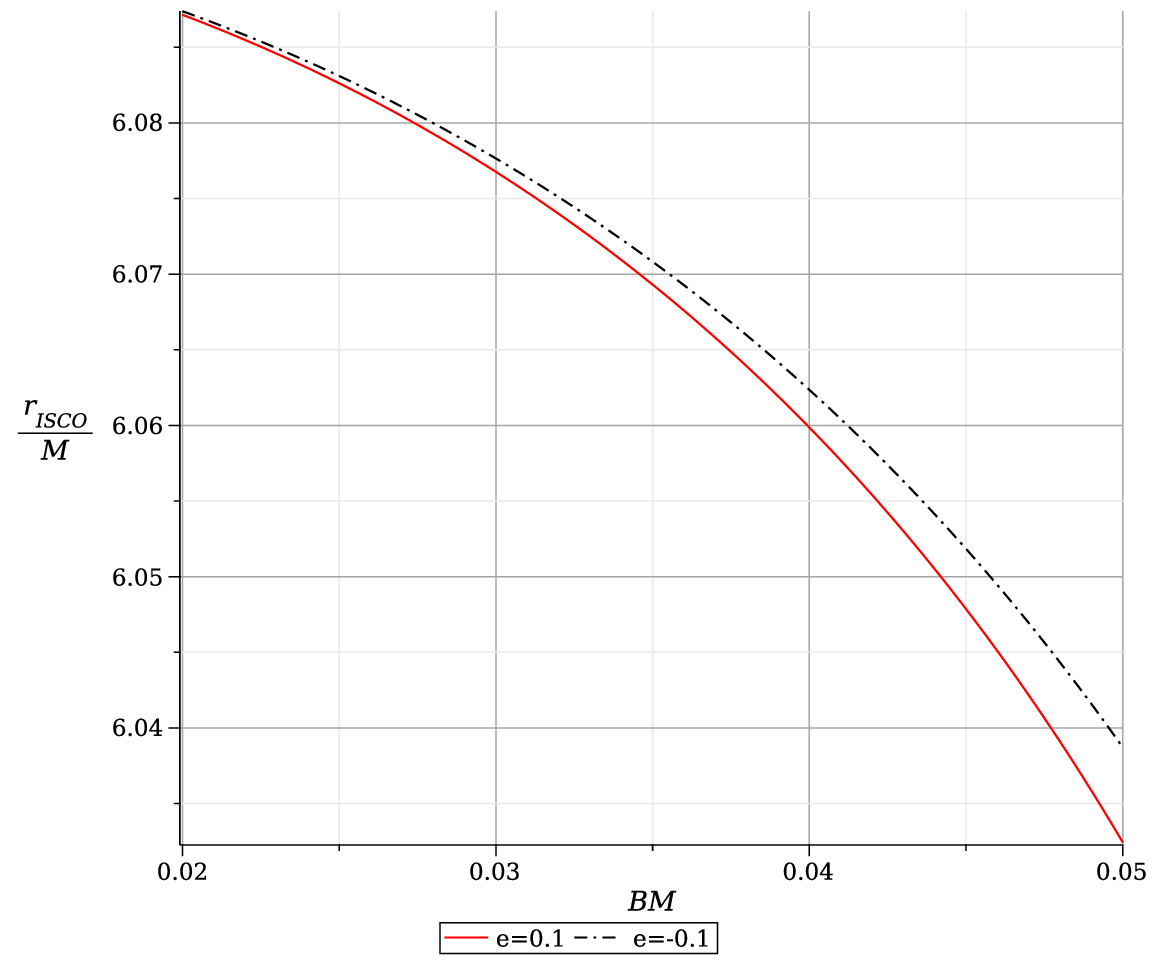}
    \caption{$C=-1$}
    \label{fig:retroIsco_cminus1}
\end{subfigure}
\begin{subfigure}{0.32\textwidth}
    \centering
    \includegraphics[width=\textwidth]{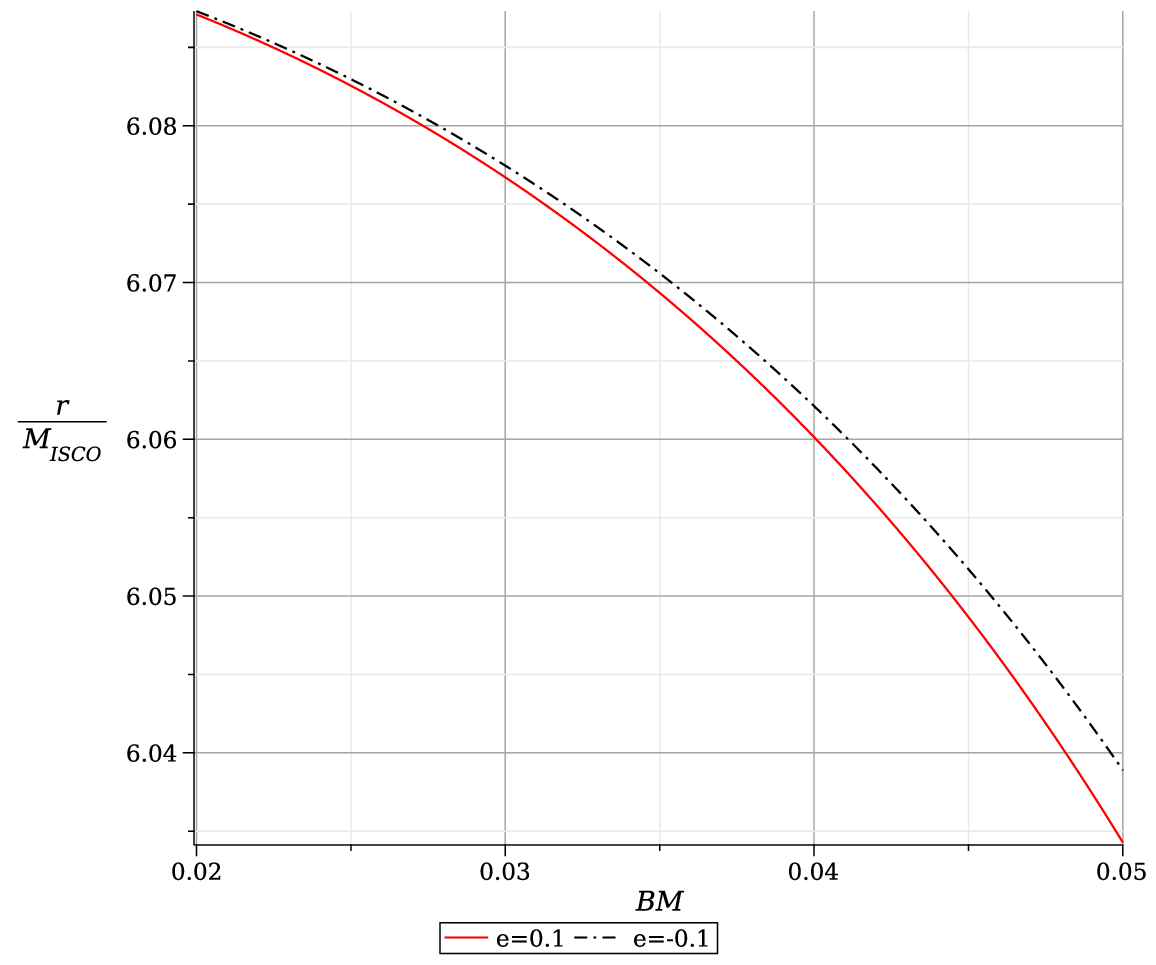}
    \caption{$C=0$}
    \label{fig:retroIsco_c0}
\end{subfigure}
\begin{subfigure}{0.32\textwidth}
    \centering
    \includegraphics[width=\textwidth]{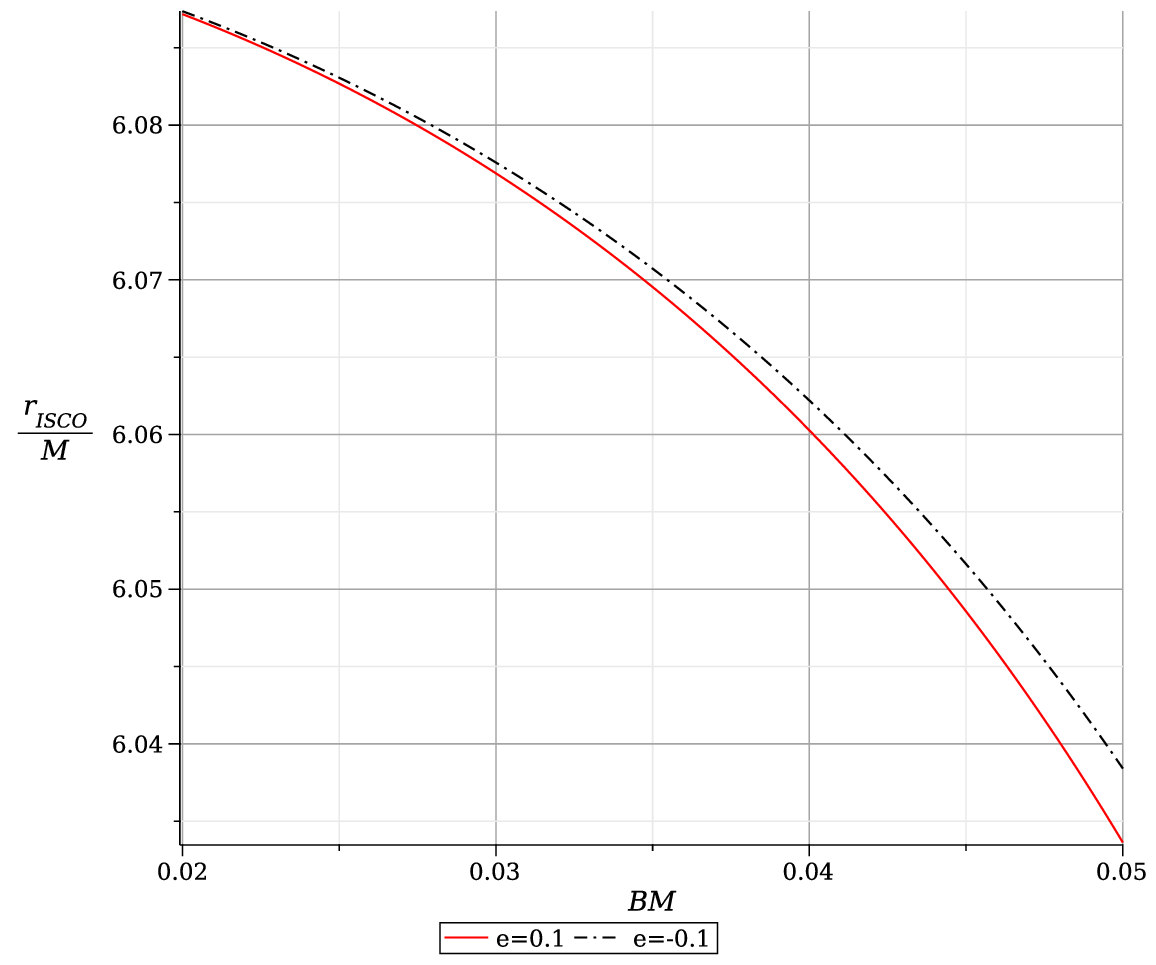}
    \caption{$C=1$}
    \label{fig:retroIsco_c1}
\end{subfigure}
 \caption{Retrograde ISCO radius $r_{\mathrm{ISCO}}$ as a function of $B$ in the weakly 
magnetized Taub--NUT spacetime with $l = 0.2$. Panels $(a)$--$(c)$ correspond to 
$C = -1, 0, +1$. Red solid curves: $e > 0$; black dash-dotted curves: $e < 0$. 
As in the prograde case, $r_{\mathrm{ISCO}}$ decreases monotonically with $B$, 
with a small charge-sign splitting; the dependence on $C$ remains comparatively weak.}
   \end{figure}
Solving the on-shell condition $g^{\mu\nu}(p_\mu - eA_\mu)(p_\nu - eA_\nu)=-\mu^2$ for
the energy (with $p_t=-E$, $p_\phi=L$) at fixed $r$ and $L$ on a circular orbit
yields an analytic relation $E(B)$:
\begin{equation}\label{eq:EofB}
E = -eA_t + \frac{1}{g^{tt}}\Bigl\{g^{t\phi}(L - eA_\phi) \mp \sqrt{(g^{t\phi})^2(L-eA_\phi)^2 - g^{tt}\bigl[g^{\phi\phi}(L-eA_\phi)^2 + \mu^2\bigr]}\Bigr\},
\end{equation}
where $g^{tt}, g^{t\phi}, g^{\phi\phi}$ denote the contravariant metric components on
the equatorial plane and $\mu$ is the particle rest mass; setting $\mu=1$ gives the
specific quantities used in the figures. The $\mp$ sign distinguishes the two roots,
selected by physical regularity. The dependence of $E$ on $B$ in the magnetized
Taub--NUT spacetime is plotted in Figs.~6 and 7.

   \begin{figure}[ht!]
	\centering
\begin{subfigure}{0.325\textwidth}
    \centering
    \includegraphics[width=\textwidth]{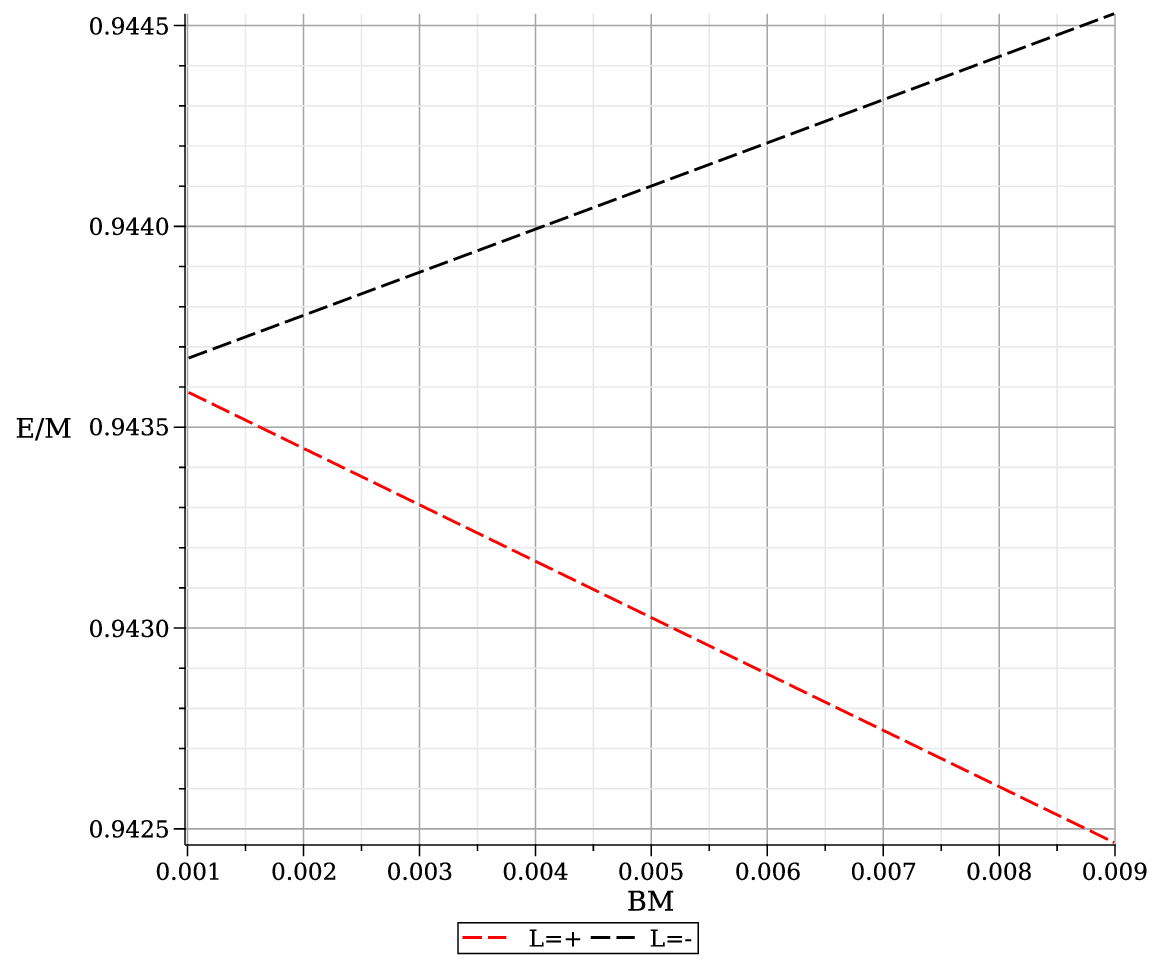}
    \caption{$C=-1$}
    \label{fig:EBposq_cminus1}
\end{subfigure}
\begin{subfigure}{0.325\textwidth}
    \centering
    \includegraphics[width=\textwidth]{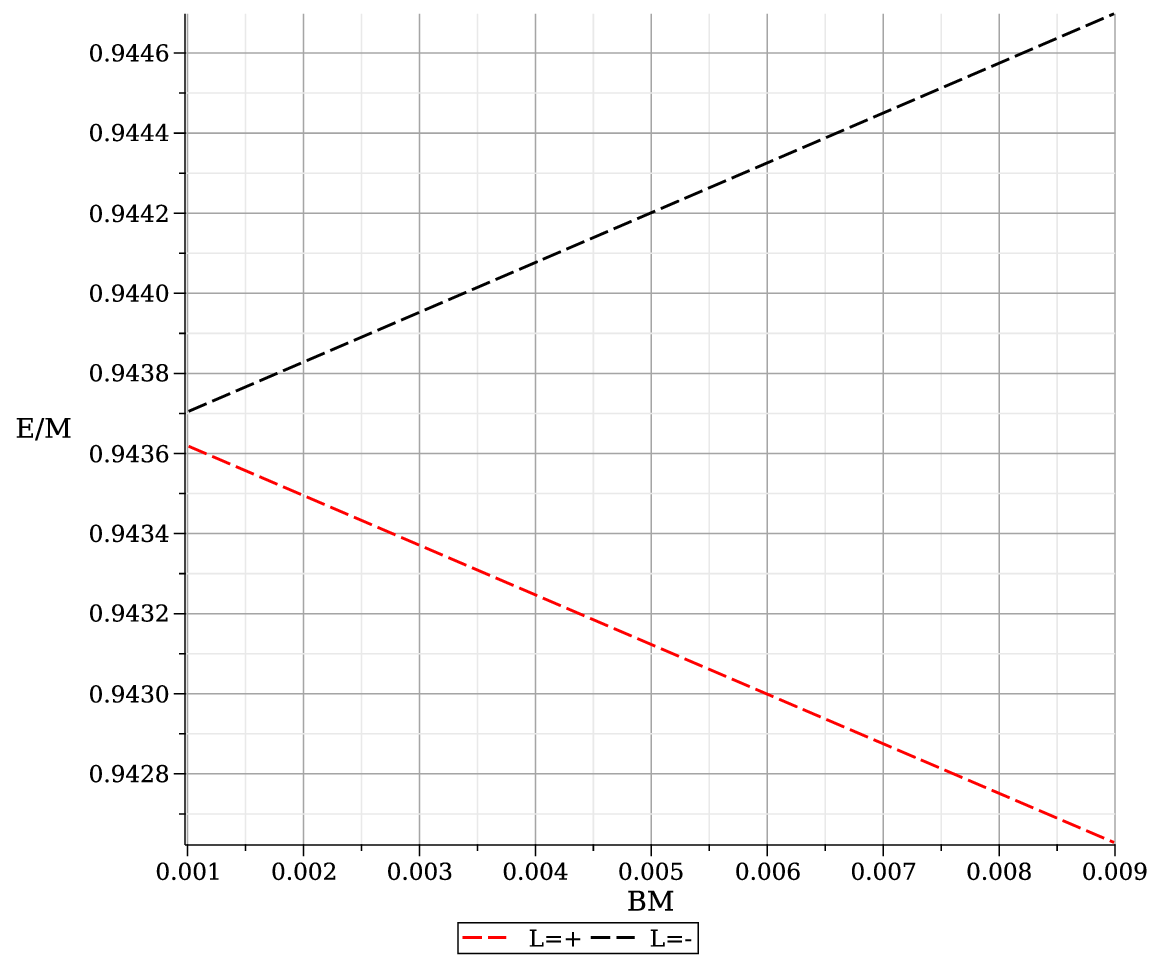}
    \caption{$C=0$}
    \label{fig:EBposq_c0}
\end{subfigure}
\begin{subfigure}{0.325\textwidth}
    \centering
    \includegraphics[width=\textwidth]{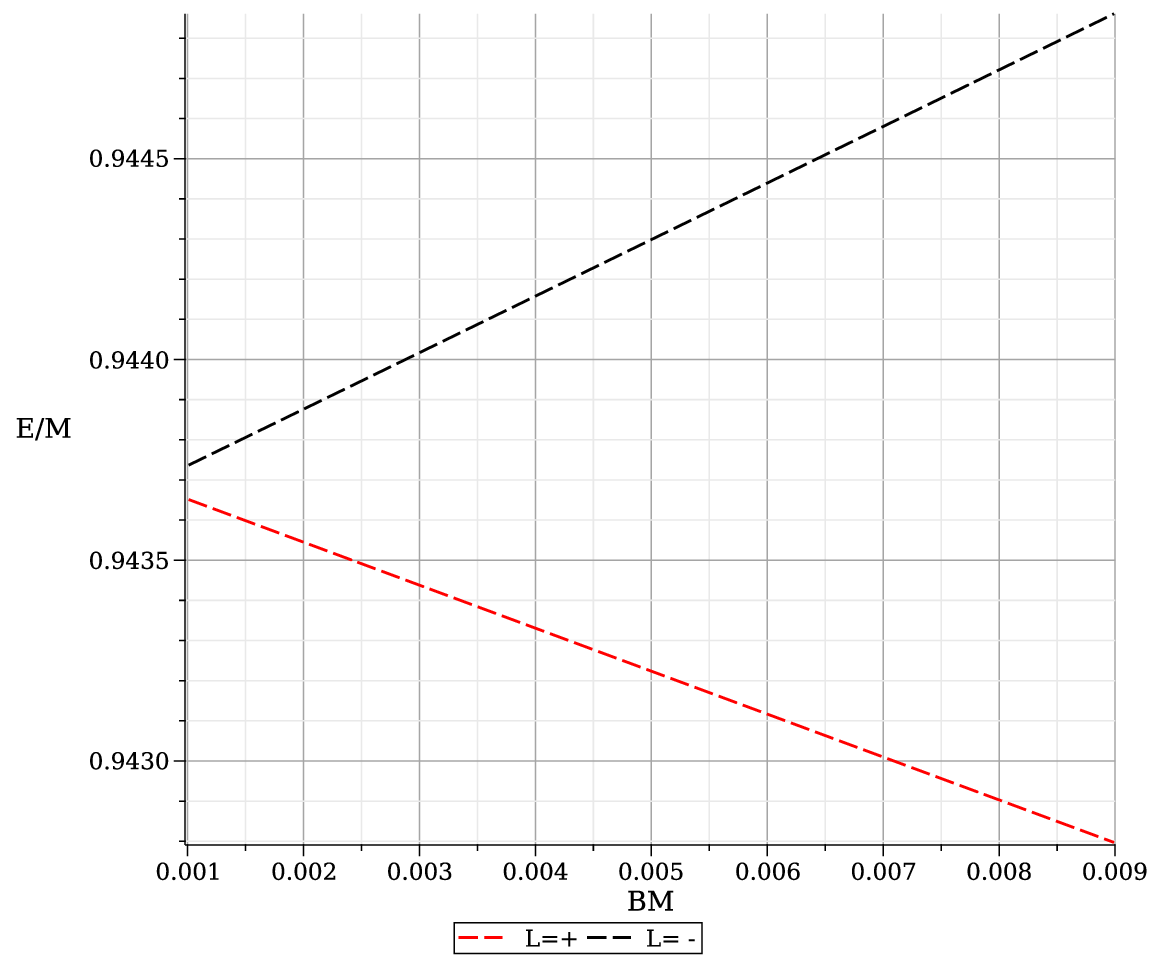}
    \caption{$C=1$}
    \label{fig:EBposq_c1}
\end{subfigure}
 \caption{Energy $(E)$ as a function of $B$ in the weakly magnetized Taub--NUT spacetime with $l = 0.2$. Panels $(a)$--$(c)$ correspond to $C = -1, 0, +1$ and $e > 0$. Red dash-dotted curves: $L > 0$; black dash-dotted curves: $L < 0$. Curve indicating the perturbative effect of the magnetic field on the stability of particle orbits}. 
   \end{figure}
   
      \begin{figure}[ht!]
	\centering
\begin{subfigure}{0.325\textwidth}
    \centering
    \includegraphics[width=\textwidth]{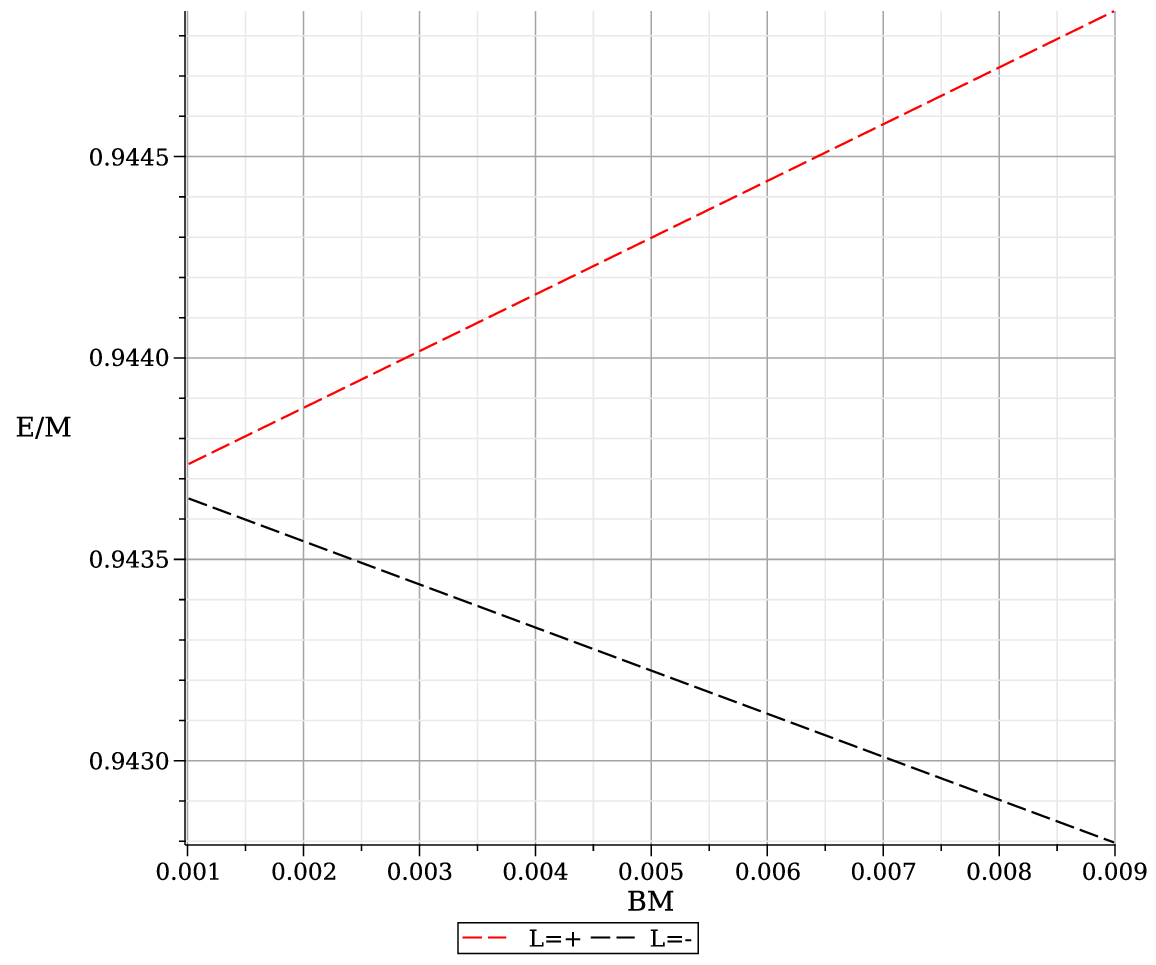}
    \caption{$C=-1$}
    \label{fig:EBnegq_cminus1}
\end{subfigure}
\begin{subfigure}{0.325\textwidth}
    \centering
    \includegraphics[width=\textwidth]{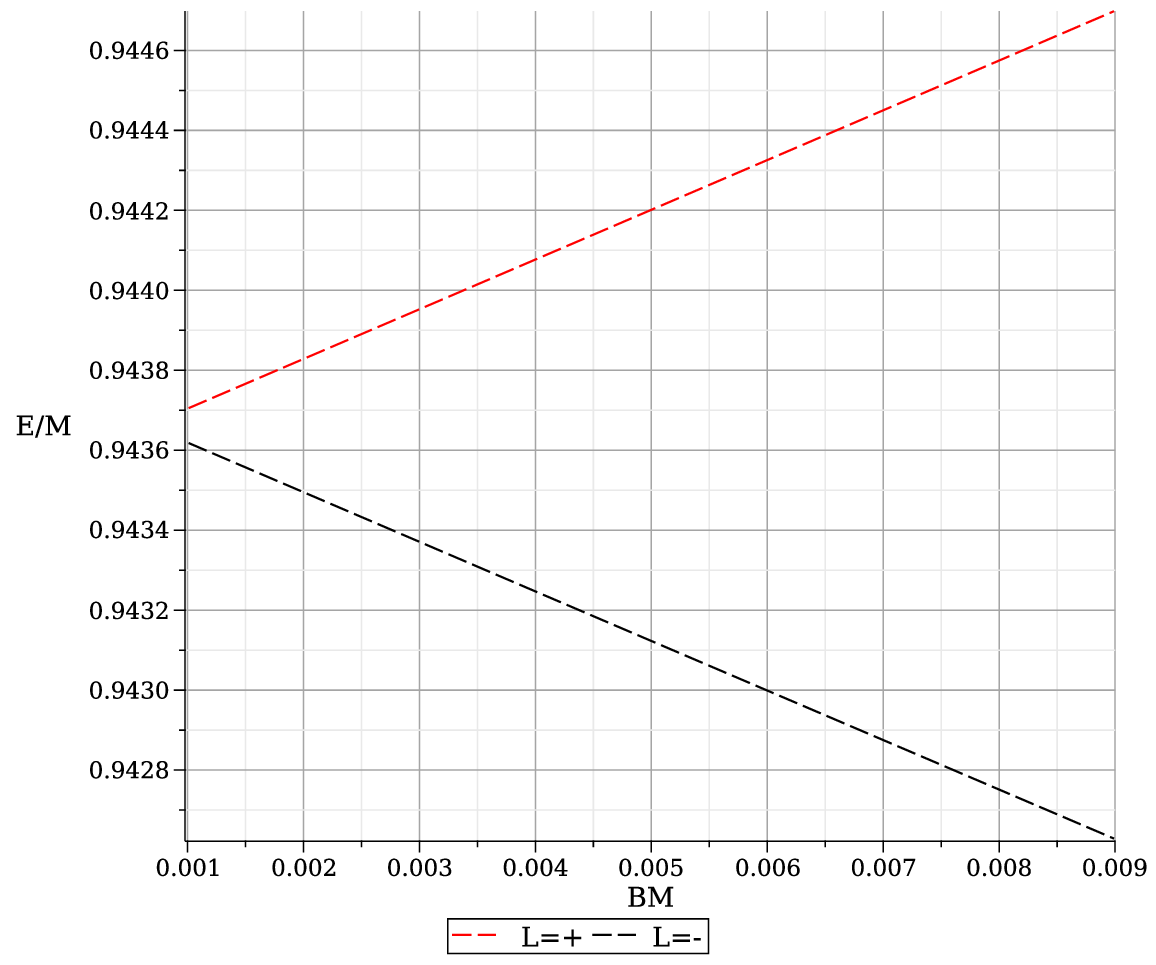}
    \caption{$C=0$}
    \label{fig:EBnegq_c0}
\end{subfigure}
\begin{subfigure}{0.32\textwidth}
    \centering
    \includegraphics[width=\textwidth]{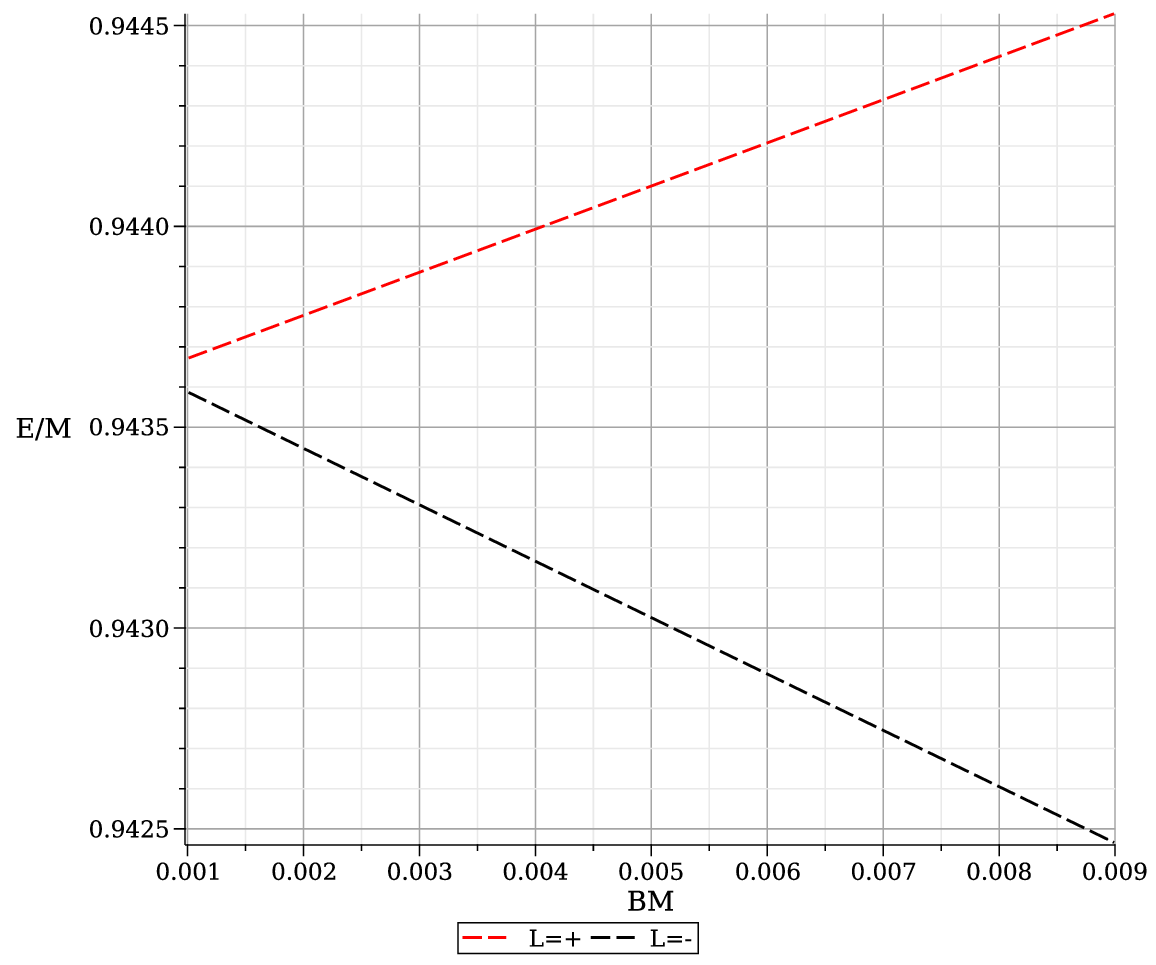}
    \caption{$C=1$}
    \label{fig:EBnegq_c1}
\end{subfigure}
 \caption{Energy $(E)$ as a function of $B$ in the weakly magnetized Taub--NUT spacetime with $l = 0.2$. Panels $(a)$--$(c)$ correspond to $C = -1, 0, +1$ and $e < 0$. Red dash-dotted curves: $L > 0$; black dash-dotted curves: $L < 0$}.
   \end{figure}

Furthermore, the relationship between angular momentum and the magnetic field can be
expressed analytically as a function $L(B)$. This is important for characterising the
dynamical stability of particle orbits in the magnetized Taub--NUT spacetime with the
Manko--Ruiz parameter, as follows:
\begin{equation}\label{eq:LofB}
L = eA_{\phi} - \frac{1}{g^{\phi\phi}}\Bigl\{g^{t\phi}(-E - eA_t) \pm \sqrt{(g^{t\phi})^2(-E-eA_t)^2 - g^{\phi\phi}\bigl[g^{tt}(-E-eA_t)^2 + \mu^2\bigr]}\Bigr\}.
\end{equation}
The dependence of the angular momentum $L$ on the magnetic field $B$ in the
magnetized Taub--NUT spacetime is plotted in Fig.~8.

      \begin{figure}[ht!]
	\centering
\begin{subfigure}{0.315\textwidth}
    \centering
    \includegraphics[width=\textwidth]{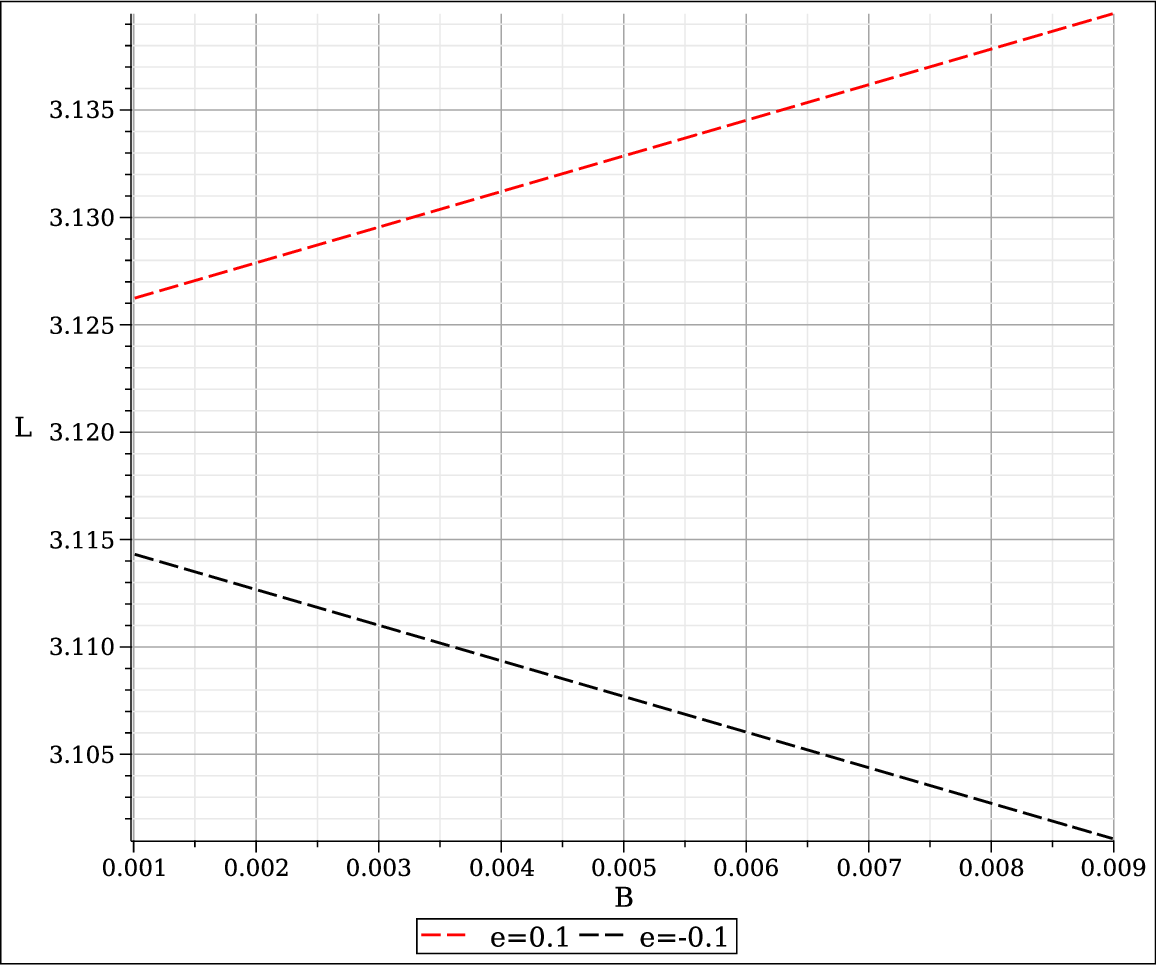}
    \caption{$C=-1$}
    \label{fig:LB_cminus1}
\end{subfigure}
\begin{subfigure}{0.315\textwidth}
    \centering
    \includegraphics[width=\textwidth]{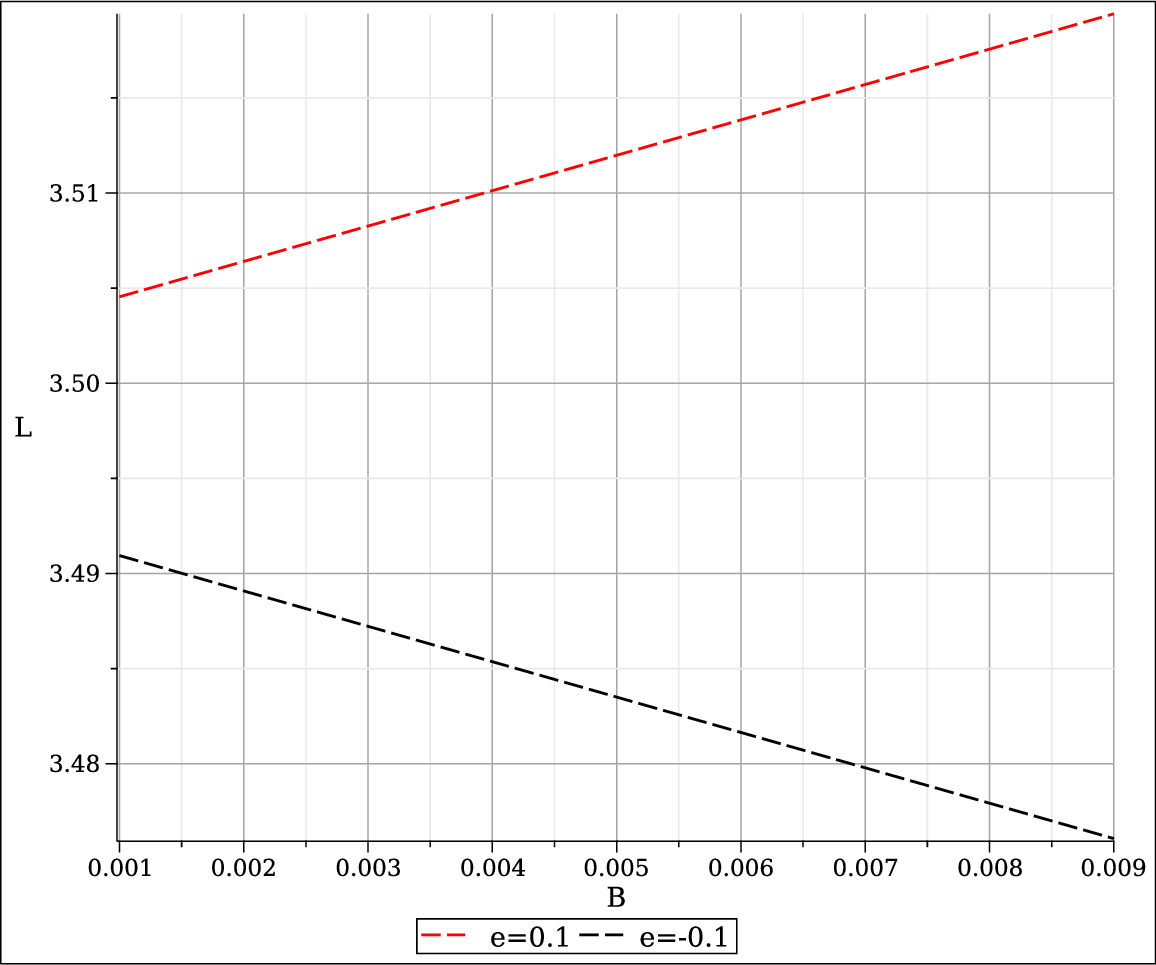}
    \caption{$C=0$}
    \label{fig:LB_c0}
\end{subfigure}
\begin{subfigure}{0.315\textwidth}
    \centering
    \includegraphics[width=\textwidth]{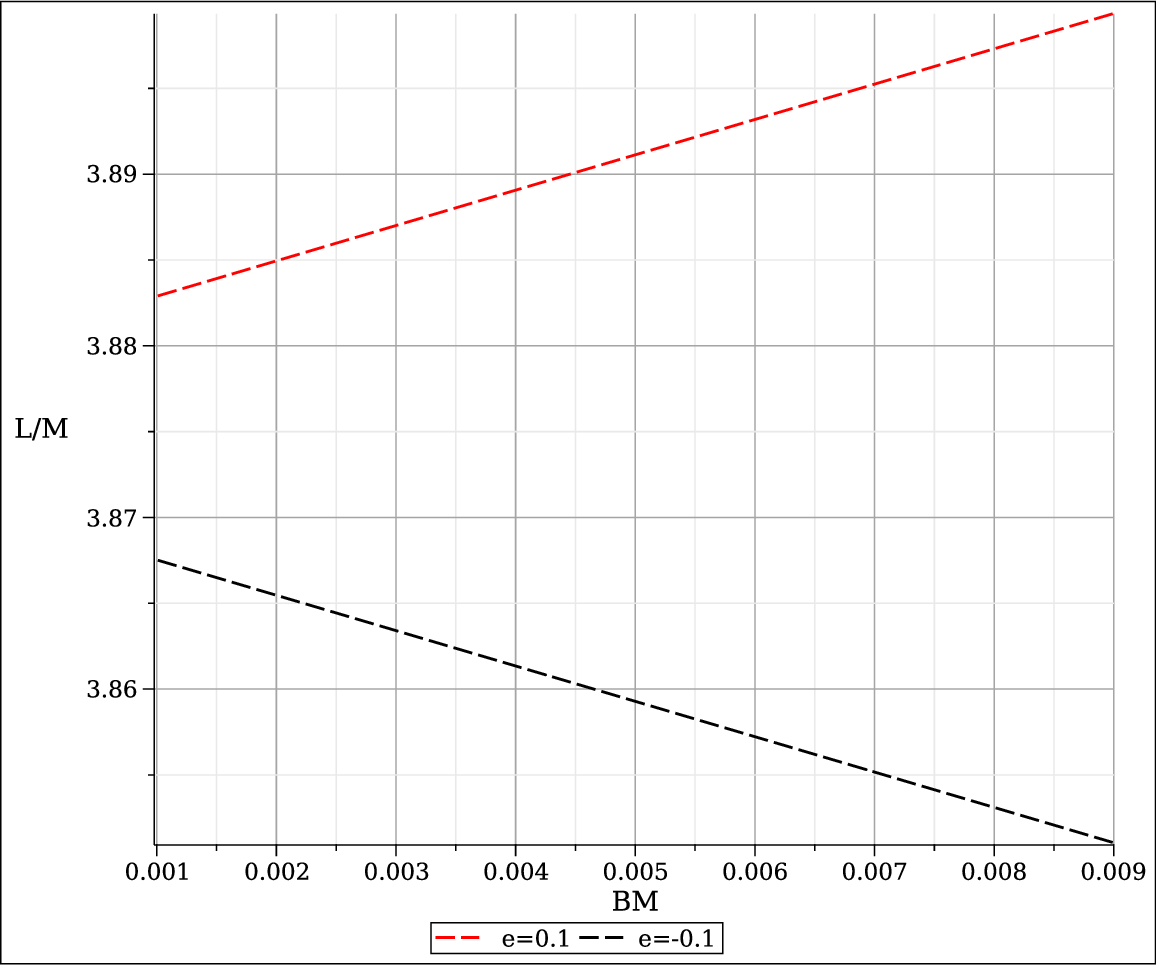}
    \caption{$C=1$}
    \label{fig:LB_c1}
\end{subfigure}
 \caption{Angular momentum  $(L)$ as a function of $B$ in the weakly magnetized Taub--NUT spacetime with $l = 0.2$. Panels $(a)$--$(c)$ correspond to $C = -1, 0, +1$ . Red dash-dotted curves: $e > 0$; black dash-dotted curves: $e < 0$}.
   \end{figure}
   Based on the plot, the normalised angular momentum $L/M$ exhibits a linear dependence
on the magnetic field parameter $BM$ for the different values of the parameter $C$.
For the $L_+$ configuration, the angular momentum increases linearly with $BM$,
whereas for the $L_-$ configuration it decreases linearly. This behaviour remains
consistent for all values $C=-1,0,1$, although a shift in the absolute values is
observed due to the influence of $C$. These results indicate that the magnetic field
contributes as a first-order perturbative correction to the angular momentum, while
clearly distinguishing the dynamical characteristics of Larmor and anti-Larmor orbits.
  
\section{Conclusion}\label{sec:conclusion}

We have analysed the circular motion of electrically charged test particles on the
imposed equatorial slice of a weakly magnetized Taub--NUT black hole with the
Manko--Ruiz parameter. Using Wald's prescription for the external magnetic field and
the full Lagrangian formulation, we showed that the equatorial slice is not, in
general, a true family of orbits, since the angular equation leaves a residual
constraint, Eq.~(\ref{eq:angconstraint}). Within this constrained setting, we derived
compact conditions for circularity and marginal stability,
Eqs.~(\ref{eq:N0})--(\ref{eq:Ndoubleprime}), and used them to identify the leading
features of the orbital dynamics.

The dominant effect is produced by the magnetic field. For both prograde and
retrograde branches, the constrained ISCO radius $r_{\mathrm{ISCO}}$ decreases
monotonically as the field strength $B$ increases, showing that the Lorentz
interaction allows stable circular motion to persist closer to the black hole.
A second robust feature is the charge-sign splitting of the ISCO curves: the
difference between $e>0$ and $e<0$ follows directly from the relative sign of the
Lorentz force and the orbital angular momentum, consistently with the leading linear
coupling $eBL$. By contrast, the Manko--Ruiz parameter $C$ affects local orbital
observables only at subleading order. Its apparent contribution to canonical
quantities such as $E$ and $L$ may partly reflect the choice of Misner-string gauge,
so a sharper separation between physical and gauge-dependent effects remains to be
established.

The present results should nevertheless be interpreted with care. They are obtained
on the imposed slice $x=\dot x=0$ without enforcing the angular constraint
self-consistently, whereas generic charged Taub--NUT orbits are expected to lie on
cones $x=x_0\neq 0$ rather than on the equatorial plane. The ISCO values reported
here therefore describe constrained motion on an imposed slice, not the true ISCO
radii of self-consistent circular orbits. In addition, we have not yet carried out
an explicit orthonormal-tetrad analysis of the ZAMO-measured electric and magnetic
fields, so the electromagnetic discussion should be regarded as based on the
coordinate components of $F_{\mu\nu}$.

Even with these limitations, the analysis provides a systematic and gauge-aware
picture of how the NUT-induced gravitomagnetic structure and an external magnetic
field jointly modify charged-particle dynamics in magnetized Taub--NUT spacetimes.
The most natural next step is to construct self-consistent conical circular orbits by
solving the angular, circularity, and marginal-stability conditions simultaneously,
and then to complement that analysis with a quantitative study of Lyapunov
exponents. Further extensions include electrically charged backgrounds in the
Reissner--Nordstr\"om--Taub--NUT family, off-equatorial bound motion, and possible
observational implications for quasi-periodic oscillations and black-hole shadow
phenomenology.

\section*{Acknowledgement}
The authors are grateful to H.~M.~Siahaan for suggesting this problem and for
invaluable discussions throughout the preparation of this manuscript.
BJ is indebted to the Faculty of Mathematics and Natural Science for financial 
support. TS is thankful to Sulawesi Barat University for the Leader 
Research Grant provided in Fiscal Year 2025.

\end{document}